\documentclass{article}

\pdfoutput=1

\usepackage{PRIMEarxiv}

\usepackage[utf8]{inputenc} 
\usepackage[T1]{fontenc}    
\usepackage{hyperref}       
\usepackage{url}            
\usepackage{booktabs}       
\usepackage{amsfonts}       
\usepackage{nicefrac}       
\usepackage{microtype}      
\usepackage{lipsum}
\usepackage{fancyhdr}       
\usepackage{graphicx}       
\graphicspath{{media/}}     

\usepackage[square,numbers]{natbib}
\usepackage{graphicx}
\usepackage{caption}
\usepackage{subcaption}
\usepackage{multirow}
\usepackage{amsmath}
\usepackage{tabu}

\title{MaxViT-UNet: Multi-Axis Attention for Medical Image Segmentation
}

\author{
  Abdul Rehman Khan\textsuperscript{1, 2}, Asifullah Khan\textsuperscript{1, 2, 3*}\\
  \textsuperscript{1} Pattern Recognition Lab, Department of Computer \& Information Sciences, Pakistan Institute of \\Engineering \& Applied Sciences, Nilore, Islamabad, 45650, Pakistan \\
  \textsuperscript{2} PIEAS Artificial Intelligence Center (PAIC), Pakistan Institute of Engineering \& Applied Sciences, \\Nilore, Islamabad, 45650, Pakistan \\
  \textsuperscript{3} Center for Mathematical Sciences, Pakistan Institute of Engineering \& Applied Sciences, Nilore, \\Islamabad, 45650, Pakistan \\
  \texttt{Corresponding Author: \textsuperscript{*}Asifullah Khan, asif@pieas.edu.pk} \\
}

\begin{document}
\maketitle

\begin{abstract}
Since their emergence, Convolutional Neural Networks (CNNs) have made significant strides in medical image analysis. However, the local nature of the convolution operator may pose a limitation for capturing global and long-range interactions in CNNs. Recently, Transformers have gained popularity in the computer vision community and also in medical image segmentation due to their ability to process global features effectively. The scalability issues of the self-attention mechanism and lack of the CNN-like inductive bias may have limited their adoption. Therefore, hybrid Vision transformers (CNN-Transformer), exploiting the advantages of both Convolution and Self-attention Mechanisms, have gained importance. In this work, we present MaxViT-UNet, a new Encoder-Decoder based UNet type hybrid vision transformer (CNN-Transformer) for medical image segmentation. The proposed Hybrid Decoder is designed to harness the power of both the convolution and self-attention mechanisms at each decoding stage with a nominal memory and computational burden. The inclusion of multi-axis self-attention, within each decoder stage, significantly enhances the discriminating capacity between the object and background regions, thereby helping in improving the segmentation efficiency. In the Hybrid Decoder, a new block is also proposed. The fusion process commences by integrating the upsampled lower-level decoder features, obtained through transpose convolution, with the skip-connection features derived from the hybrid encoder. Subsequently, the fused features undergo refinement through the utilization of a multi-axis attention mechanism. The proposed decoder block is repeated multiple times to segment the nuclei regions progressively. Experimental results on MoNuSeg18 and MoNuSAC20 datasets demonstrate the effectiveness of the proposed technique. Our MaxViT-UNet outperformed the previous CNN-based (UNet) and Transformer-based (Swin- UNet) techniques by a considerable margin on both of the standard datasets. The following \href{https://github.com/PRLAB21/MaxViT-UNet}{github} (https://github.com/PRLAB21/MaxViT-UNet) contains the implementation and trained weights.
\end{abstract}


\section{Introduction}\label{Introduction}
In medical image analysis, particular structures or regions of interest are identified and delineated from medical images; image segmentation plays a crucial role in this process \cite{kumar2019multi}. In particular, nuclei segmentation—which entails determining the boundaries of nuclei cells in microscopic histopathological images—is an essential task \cite{kumar2019multi}. In order to increase the precision and dependability of medical diagnosis and treatment, nucleus segmentation is essential. Because it makes exact nuclei detection and quantification possible, it can help with the creation of individualized treatment programs and enhance patient results.

Deep learning algorithms have shown exceptional performance in a variety of applications \cite{tayyab2022survey, sohail2023deep, khan2023survey_covid, Rauf2023-tc, khan2008machine, naveed2012gpcr, khan2017random, chouhan2021deep, majid2006combination, zahoor2022new}, particularly image segmentation \cite{long2015fully, ronneberger2015u, cao2022swin, rauf2023attention, ali2022channel, aziz2020channel, sohail2021mitotic, khan2023segmentation, khan2023survey, zhang2022multi, vu2019dense, liu2023mestrans, khan2023recent}, in recent years. They have shown to be quite successful in completing this challenging assignment accurately and quickly. Medical image analysis greatly benefits from Convolutional Neural Networks' ability to automatically capture low-level to high-level properties hierarchically from a dataset \cite{khan2020survey}. In particular, deep CNNs are being used more often for medical image segmentation. Deep learning models such as Fully Convolutional Neural Network (FCNN) \cite{long2015fully} and UNet-like encoder-decoder architectures have emerged as powerful tools in medical image segmentation, outperforming older methods \cite{ronneberger2015u, ibtehaz2020multiresunet, zhou2019unet++, oktay2018attention, cciccek20163d, cao2022swin}. Such networks extract deep features and combine high-resolution information using a combination of convolutional and down-sampling layers in the encoder and up-sampling layers in the decoder to provide pixel-level semantic predictions. They can extract features and understand complicated patterns from medical images, making them ideal for medical image processing tasks like segmentation.

More recently, vision transformers (ViTs) \cite{dosovitskiy2021an, chen2021transunet, valanarasu2021medical, cao2022swin, wang2022uctransnet} have also emerged as a powerful tool for the medical domain and have shown impressive results on a variety of segmentation tasks in medical imaging. The idea of self-attention in ViTs helps focus on relevant image regions while suppressing irrelevant features, and significantly improves the segmentation performance for medical images. It helps in effectively extracting the most important features in an image and capturing the long-range dependencies between them \cite{dosovitskiy2021an}. Swin-UNet \cite{cao2022swin} has updated the Swin Transformer to segment medical images by presenting a decoder based on the shifting window attention mechanism. Their architecture is entirely transformer-based and lacks the inductive bias of convolutions. Hybrid approaches \cite{zhang2021transfuse, li2023attransunet, guo2022cmt, li2022next, yao2022transclaw, ji2021multi, zhou2021nnformer} try to tackle this problem by using convolutions along with self-attention in their encoder but either they suffer from quadratic nature of self-attention or use convolution-based decoders. To the best of our knowledge, Hybrid Decoder has never gained much attention for medical image segmentation tasks before. Our proposed idea utilizes a hybrid technique in both encoder and decoder and uses multi-axis attention with linear time complexity with respect to image dimensions.

Inspired by the results of Multi-Axis self-attention (Max-SA) \cite{tu2022maxvit}, we have used its potential for medical image segmentation to develop a novel architecture dubbed MaxViT-UNet, which uses a UNet-style framework. We also presented a new hybrid ViT-based decoder block and a new fusion module. By adding Max-SA, we improved the multi-head self-attention mechanism, allowing for the computationally efficient extraction of local and global-level information. The MaxViT-UNet encoder and decoder's hybrid design allows for the creation of contextually rich features at higher levels as well as noise-free features at lower levels, which is critical for accurate medical image segmentation. The primary results of our methodology are as follows:

\begin{enumerate}
\item \textbf{MaxViT-UNet}: A comprehensive hybrid architecture for medical image segmentation, comprised of a MaxViT-based Hybrid Encoder and a Proposed Hybrid Decoder. Both employ hybrid CNN-Transformer blocks and skip connections at all scales for effective feature processing.
\item \textbf{Hybrid Decoder Block}: This two-step module (1) merges up-sampled features from higher semantic levels with encoder features from high spatial levels using simple concatenation, and (2) fuses the concatenated information through efficient CNN-Transformer blocks.
\item \textbf{Parameter-Efficient Design}: The decoder block's repetitive structure promotes parameter efficiency and computational lightness without sacrificing segmentation performance. Local and global attention throughout the decoder aids in discarding irrelevant features for high-quality segmentation.
\item \textbf{Experimental Validation}: MaxViT-UNet's effectiveness is demonstrated across multiple datasets. Ablation studies further support the promising use of the Hybrid Decoder for medical image segmentation.
\end{enumerate}

The paper begins by reviewing recent medical image segmentation methods, categorized by their reliance on convolutional neural networks (CNNs), transformers, or hybrid approaches (Section \ref{Related Works}). Section \ref{Proposed Methodology for MaxViT-UNet Framework} delves into the proposed MaxViT-UNet architecture, detailing its encoder and decoder components, hybrid blocks, and key innovations. Section \ref{Experiments and Results} then describes the experimental setup, including datasets, evaluation measures, and accomplished outcomes, followed by a discussion of the findings and their consequences. The final section of the study discusses probable future directions.

\section{Related Works}\label{Related Works}

The conventional approaches for medical image segmentation primarily relied on morphological operations (opening, closing, dilation, and erosion), or contour, color, and watershed-based techniques and traditional machine learning \cite{yang2006nuclei, veta2013automatic, tsai2003shape, held1997markov}. These approaches do not generalize well and suffer from different sources of variations in medical images such as variation in nuclei shape across various organs and tissue types, variation in color across crowded and sparse nuclei, variation in imaging equipment and hospitals/clinics protocol \cite{kumar2019multi}.

\subsection{CNN Based Techniques}\label{CNN Based Techniques}

One of the first approaches to deep learning-based medical image segmentation used FCN (fully convolutional network) \cite{long2015fully}. Despite outperforming conventional techniques, FCN's pooling procedure resulted in the loss of texture and edge information, which are required for segmentation. Therefore, Ronneberger et al. \cite{ronneberger2015u} proposed an encoder-decoder structure called UNet by improving the idea of FCN. To mitigate for semantic loss, the U-shaped architecture connected the encoder and decoder at various stages via skip-connections. UNet's simple and unique architecture gave exceptional performance, prompting the creation of other variations in various medical image domains for segmentation purposes. MultiResUNet \cite{ibtehaz2020multiresunet} replaced skip-connections with residual paths to extract semantic information at multiple scales. M-Net \cite{fu2018joint} captured multi-level semantic details by injecting rich multi-scale input features into different layers and processing them through a couple of downsampling and upsampling layers. UNet++ \cite{zhou2019unet++} proposed a new variant of UNet that incorporates dense connections nested together to effectively represent fine-grained object information. DenseRes-Unet \cite{kiran2022denseres} used a dense bottleneck layer in Unet architecture for nuclei segmentation. With the help of channel-wise stitching in the encoder and skip connections in the decoder, AlexSegNet \cite{singha2023alexsegnet} uses an encoder-decoder framework based on the AlexNet architecture to combine low-level and high-level features. Recently, the idea of an attention mechanism has been applied to enhancing the segmentation performance of medical systems: AttentionUNet \cite{oktay2018attention} improved the segmentation performance of medical images using soft attention by introducing an attention-gate module. The CA-Net \cite{gu2020net} merged the Spital, Channel, and Scale attentions into one comprehensive attention mechanism. Attention Assisted UNet \cite{wang2019sclerasegnet} also improved the attention mechanism for accurate segmentation of sclera images. NucleiSegNet \cite{lal2021nucleisegnet} utilized attention in the decoding stage. Cell-Net \cite{shi2022fine} uses multiscale and dilated convolutions to capture both global and local characteristics. Some notable work for end-to-end 3D medical image segmentation includes 3D UNet \cite{cciccek20163d}, which replaced 2D convolution with 3D convolution. V-net \cite{milletari2016v} also improved the UNet with the help of 3D convolution and proposed a dice loss for better segmentation masks. Recently, the idea of CB-CNN (Channel Boosted-CNN) has also emerged for medical image segmentation tasks \cite{ali2022channel, aziz2020channel, sohail2021mitotic, ali2023cb}, where diverse feature spaces from multiple encoders are fused together to improve the quality of segmentation models.

\subsection{ViT Based Techniques}\label{ViT Based Techniques}

The CNN-based U-shaped networks are effective, but the convolution operation captures only the local information and discards global information. To prevent misclassification in segmentation, it is crucial to learn the long-range dependency between background and mask pixels. However, constructing deeper networks or utilizing larger convolution kernels to capture long-range relationships results in an explosion of parameters, making the training process more expensive. To solve these challenges, Dosovitskiy et al. \cite{dosovitskiy2021an} developed Vision Transformers (ViT), which have a multi-headed self-attention mechanism capable of capturing long-range dependencies in computer vision tasks. After the success of ViT in natural images and large datasets, medical image processing has also evolved with transformer-based techniques. One of the first techniques that combined transformer and UNet is TransUNet \cite{chen2021transunet}. Later, to efficiently handle smaller-sized medical image datasets, Medt \cite{valanarasu2021medical} improved the self-attention mechanism using a gated axial attention module. Swin-Unet \cite{cao2022swin}, a pure transformer architecture, adopted the Swin Transformer into a U-shaped encoder-decoder segmentation framework to capture local as well as global semantic features in a hierarchical fashion. UCTransNet \cite{wang2022uctransnet} replaced the skip-connection with the channel transformer (CTrans) module. Karimi et al. \cite{karimi2021convolution} used self-attention between surrounding image patches to change the MHSA mechanism of vision transformers. Despite excelling at several image segmentation tasks, ViTs suffer from the problem of computational overload.

\subsection{Hybrid Based Techniques}\label{Hybrid Based Techniques}

The Transformer-based design outperforms CNN in collecting long-range dependencies, but it suffers from a lack of interaction with surrounding feature information due to image split into fixed-sized patches. Surpassing the existing performance of medical image segmentation systems is challenging to achieve with transformer-based or CNN-based UNet designs. To overcome this challenge and enhance segmentation performance, researchers have combined CNNs with Transformers by utilizing the convolution operation along with self-attention operation to create a lightweight hybrid image segmentation framework \cite{zhang2021transfuse,li2023attransunet,guo2022cmt,li2022next}. Chen et al. \cite{chen2021transunet} proposed a strong encoder by combining a Transformer with CNN for segmenting 2D medical images. Claw UNet \cite{yao2022transclaw} used the complementarity of Transformer and CNN to create hybrid blocks in the encoder for multi-organ dataset segmentation. Multi-Compound Transformer \cite{ji2021multi} achieved cutting-edge performance in six different benchmarks by integrating semantic information of hierarchical scales into a unified framework. Zhou et al. \cite{zhou2021nnformer} applied CNN and transformer blocks in a crosswise manner to achieve better performance.

The above-mentioned hybrid approaches continue to suffer from the quadratic nature of the self-attention process, preventing them from correctly combining CNNs' local feature extraction capabilities with Transformers' global feature extraction capability. In the proposed technique, we efficiently interleaved convolution and self-attention at each stage, and the linear character of the multi-axis attention mechanism used makes our technique quick and robust for medical image segmentation.

\section{Proposed Methodology for MaxViT-UNet Framework}\label{Proposed Methodology for MaxViT-UNet Framework}

\subsection{Architecture Overview of MaxViT-UNet Framework}\label{Architecture Overview of MaxViT-UNet Framework}

\begin{figure}[ht!]
  \centering
  \includegraphics[width=0.8\linewidth]{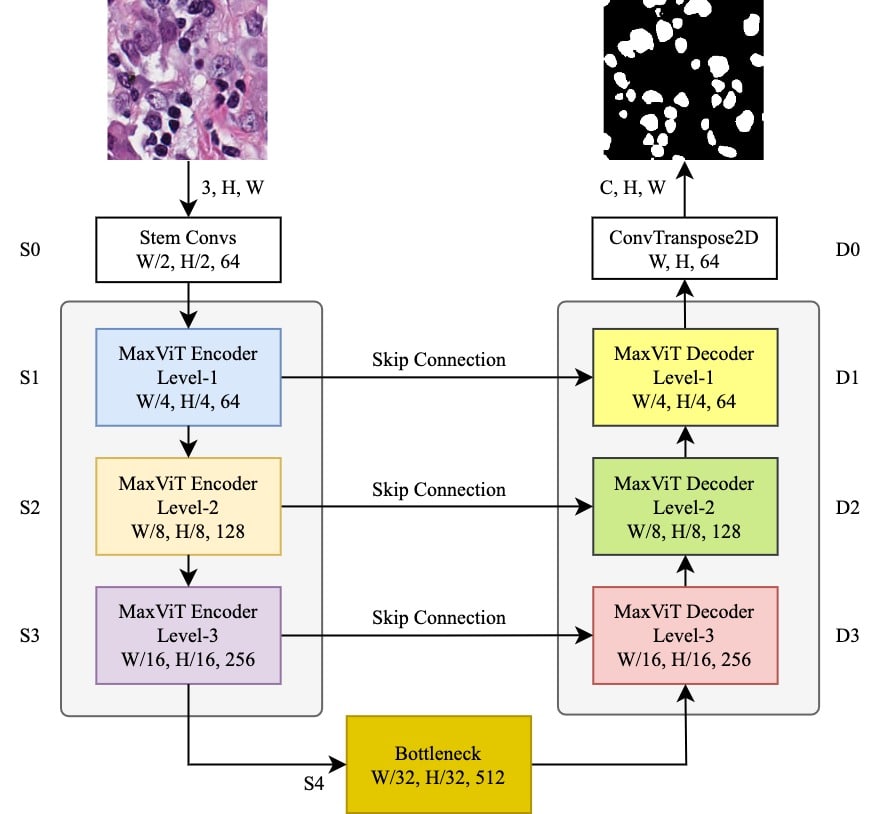}
  \caption{Encoder-Decoder architecture of the Proposed MaxViT-UNet. The encoder generates hierarchical features at four scales. The proposed decoder first upscale the bottom-level features, merges them with skip-connection features, and applies the MaxViT-based hybrid feature processing blocks a couple of times to produce an output mask image for the "C" number of classes.}
  \label{fig:maxvit-unet-architecture}
\end{figure}

The proposed MaxViT-UNet includes an encoder, a bottleneck layer, a proposed decoder, and encoder to decoder skip connections. Figure \ref{fig:maxvit-unet-architecture} presents the complete architectural details of the proposed methodology. Throughout our encoder-decoder architecture, we utilized the identical MaxViT block structure, consisting of a parameter-efficient MBConv \cite{sandler2018mobilenetv2} and scalable Max-SA mechanisms \cite{tu2022maxvit}. The stem stage of the encoder (\(S0\)) downsamples the input image of shape \(C \times H \times W\) into \(64 \times \frac{H}{4} \times \frac{W}{4}\) using Conv3$\times$3 layers. The input sequentially passes through four encoder stages \(S1\) to \(S4\). Each encoding stage doubles the feature channels (\(64, 128, 256, 512\)) while havling the spatial size (\(\frac{1}{4}, \frac{1}{8}, \frac{1}{16}, \frac{1}{32}\)), creating hierarchical features like UNet. The first MBConv block in each stage is responsible for doubling the input channels using the Conv1$\times$1 layer and halving the spatial size using the Depthwise-Conv3$\times$3 layer. The last encoder layer, also called bottleneck, contains contextually rich features and provides a bridge from encoder to decoder.

The symmetric nature of the proposed hybrid decoder, comprising the Hybrid MaxViT blocks, is inspired by UNet \cite{ronneberger2015u}. The decoder is made up of three stages, $D1$ to $D3$, matching with $S1$ to $S3$ stages of the encoder, respectively. Spatial information is lost during the encoder's downsampling process. To overcome this information, at each stage of the decoder, contextually rich features from the lower decoder stage are concatenated with locality-rich features of the encoder transferred through skip-connection. In contrast to the Conv3$\times$3 layer in the encoder that shrinks the spatial size, a Transpose Convolution layer is used for up-sampling the feature maps of the previous stage. The concatenated features are transformed through a couple of hybrid MaxViT blocks before passing to the next stage. MaxSA processing helps reduce the noise information and simultaneously models the local-global differences between background and nuclei pixels. After the decoder’s feature processing stages, the feature maps of shape \(64 \times \frac{H}{4} \times \frac{W}{4}\) are up-sampled four times to make output mask have same dimensions as that of the input image and true mask \(H \times W\). The last convolution layer reduces the channels from 64 to C (number of classes) to generate the pixel-level segmentation probabilities for each class. The proposed MaxViT-UNet, though hybrid in nature, consists of only 24.72 million parameters, lighter than UNet with 29.06 million parameters and Swin-UNet with 27.29 million parameters. In terms of computation, the proposed MaxViT-UNet takes 7.51 GFlops as compared to UNet and Swin-UNet which take 50.64 and 11.31 GFlops. Table \ref{table:maxvit-unet-architecture} summarizes the architectural configurations of the MaxViT-UNet.

\begin{table}[ht!]
  \centering
  \caption{Configuration of the Proposed MaxViT-UNet architecture.}\label{table:maxvit-unet-architecture}
  \begin{tabular*}{\textwidth}{@{\extracolsep\fill}ccc}
    \toprule%
      \textbf{Encoder Level} & \textbf{Ouput Size} & \textbf{MaxViT Encoder} \\
    \midrule
      Stem & (64, 128, 128) & $\begin{tabular}{c}
      Conv(k=3, s=2) \\
      Conv(k=3, s=1)
      \end{tabular}$ \\
    \midrule
      S1 & (64, 64, 64) & $\left\{\begin{tabular}{c}
      MBConv(E=4, R=4) \\
      Window-Rel-MSA(P=8, H=2) \\
      Grid-Rel-MSA(G=8, H=2)
      \end{tabular}\right\} \times 2 $ \\
    \midrule
      S2 & (128, 32, 32) & $\left\{\begin{tabular}{c}
      MBConv(E=4, R=4) \\
      Window-Rel-MSA(P=8, H=2) \\
      Grid-Rel-MSA(G=8, H=2)
      \end{tabular}\right\} \times 2 $ \\
    \midrule
      S3 & (256, 16, 16) & $\left\{\begin{tabular}{c}
      MBConv(E=4, R=4) \\
      Window-Rel-MSA(P=8, H=2) \\
      Grid-Rel-MSA(G=8, H=2)
      \end{tabular}\right\} \times 2/5 $ \\
    \midrule
      S4 & (512, 8, 8) & $\left\{\begin{tabular}{c}
      MBConv(E=4, R=4) \\
      Window-Rel-MSA(P=8, H=2) \\
      Grid-Rel-MSA(G=8, H=2)
      \end{tabular}\right\} \times 2 $ \\
    \midrule
      \textbf{Decoder Level} & \textbf{Ouput Size} & \textbf{Hybrid Decoder} \\
    \midrule
      \multirow{2}{*}{D1} & \multirow{2}{*}{(64, 64, 64)} & ConvTranspose(k=2, s=2) \\
      & & $\left\{\begin{tabular}{c}
      MBConv(E=4, R=4) \\
      Window-Rel-MSA(P=8, H=2) \\
      Grid-Rel-MSA(G=8, H=2)
      \end{tabular}\right\} \times 2 $ \\
    \midrule
      \multirow{2}{*}{D2} & \multirow{2}{*}{(128, 32, 32)} & ConvTranspose(k=2, s=2) \\
      & & $\left\{\begin{tabular}{c}
      MBConv(E=4, R=4) \\
      Window-Rel-MSA(P=8, H=2) \\
      Grid-Rel-MSA(G=8, H=2)
      \end{tabular}\right\} \times 2 $ \\
    \midrule
      \multirow{2}{*}{D3} & \multirow{2}{*}{(256, 16, 16)} & ConvTranspose(k=2, s=2) \\
      & & $\left\{\begin{tabular}{c}
      MBConv(E=4, R=4) \\
      Window-Rel-MSA(P=8, H=2) \\
      Grid-Rel-MSA(G=8, H=2)
      \end{tabular}\right\} \times 2 $ \\
    \hline
  \end{tabular*}
\end{table}

\subsection{MaxViT Block}\label{MaxViT Block}

The hybrid MaxViT-block effectively blends the multi-axis attention (MaxSA) mechanism with convolution, as shown in Figure \ref{fig:maxvit-unet-decoder-block}. It is based on the observation that convolution complements transformer attention by improving the generalization and the training speed of the network \cite{xiao2021early}. To this end, MBConv sub-block \cite{sandler2018mobilenetv2}, containing squeeze-and-excitation (SE) \cite{hu2018squeeze} attention, is used for feature processing before applying the multi-axis attention (MaxSA). Another benefit of the MBConv layer is that it eliminates the need for explicit positional encoding layers by acting as conditional position encoding (CPE) \cite{chu2023conditional} using depth-wise convolutions. In MBConv the expansion for the inverted bottleneck layer was set to 4 and the shrink rate for the squeeze-excitation layer was set to 0.25. After the MBConv layer, the block and grid self-attentions are stacked sequentially to model the local and global feature interactions simultaneously in a single block. Following the good design practices \cite{dosovitskiy2021an,cao2022swin}, the MaxViT block contains LayerNorm \cite{ba2016layer}, Feed-Forward Networks (FFNs) \cite{dosovitskiy2021an,cao2022swin}, and skip-connections in MBConv, block and grid attentions sub-blocks.

Let \textbf{z} represents input feature tensor, the MBConv block without downsampling is given as:
\begin{align}
\textbf{z} &= \textbf{z} + \texttt{PROJ}(\texttt{SE}(\texttt{DWCONV}(\texttt{CONV}(\texttt{BN}(\textbf{z})))))
\end{align}

where \texttt{BN} represents BatchNorm layer \cite{ioffe2015batch}, \texttt{CONV} is expanding layer consisting of Conv1$\times$1, BatchNorm and GELU \cite{hendrycks2016gaussian} activation function. \texttt{DWCONV} is processing layer consisting of Depthswise-Conv3$\times$3, BatchNorm and GELU. \texttt{SE} represents Squeeze-Excitation layer \cite{hu2018squeeze}, while \texttt{PROJ} reduces the number of channels using Conv1$\times$1.

In each stage, the first MBConv downsample the input \textbf{z} using Depthswise-Conv3$\times$3 with a stride of 2, while the residual connection consists of pooling and channel projection layers:
\begin{align}
\textbf{z} &= \texttt{PROJ}(\texttt{MAXPOOL}(\textbf{z})) + \texttt{PROJ}(\texttt{SE}(\texttt{DWCONV}\downarrow(\texttt{CONV}(\texttt{BN}(\textbf{z})))))
\end{align}

\subsection{Max-SA: Multi-Axis Self-Attention}\label{Max-SA: Multi-Axis Self-Attention}

The self-attention introduced by transformer \cite{vaswani2017attention} and utilized by vision transformer \cite{dosovitskiy2021an} fall into the category of dense attention mechanism due to its quadratic complexity. Considering the effectiveness of sparse approaches for self-attention \cite{tu2022maxim,zhao2021improved}, Tu et al. \cite{tu2022maxvit} presented a successful and scalable self-attention module called multi-axis self-attention, which decomposes the original self-attention into sparse forms. (1) Window Attention for blocked local feature extraction and (2) Grid Attention for dilated global feature processing. Max-SA provides linear complexity without losing locality information.

The blocked local or window attention follows the idea of Swin Transformer \cite{cao2022swin}. Let \textbf{z} represents a feature tensor of shape \(C \times H \times W\). The window partition layer reshapes \(C \times H \times W\) into shape (\(N, P \times P, C\)), where \(N=\frac{H}{P} \times \frac{W}{P}\) represents the total number of non-overlapping local windows, each of spatial shape \(P \times P\) and channel dimension \(C\). Each local window is passed through standard multi-head self-attention (MHSA) to model the local interactions. Finally, the window reverse layer reshapes \(C\) back to \(C \times H \times W\).

In order to model global interactions, Max-SA incorporates grid attention, a simple and effective way of obtaining global relations in a sparse manner. The grid partition layer reshapes \textbf{z} into shape (\(G \times G, N, C\)), to obtain \(G \times G\) number of global windows, each having dynamic spatial size represented with \(N=\frac{H}{G} \times \frac{W}{G}\), and channel dimension \(C\). To represent the local interactions, each local window is subjected to typical multi-head self-attention (MHSA). Finally, the window reverse layer reshapes \textbf{z} back to \(C \times H \times W\). The utilization of grid-attention on the decomposed grid axis enables the global mixing of spatial tokens through dilated operations. Linear complexity with respect to spatial size is guaranteed by the Max-SA technique, which maintains consistent window and grid sizes.

The location equivariance inductive bias of CNNs is well-known, and it is a feature absent from standard self-attention mechanisms \cite{dosovitskiy2021an,han2021transformer}. To address this issue, Max-SA attention blocks embraced the pre-normalized relative self-attention \cite{dai2021coatnet} A learnable relative bias is added to the attention weights by the relative self-attention mechanism \cite{dai2021coatnet,shaw2018self,jiang2021transgan,cao2022swin}, which has been shown to enhance the self-attention mechanism.

Max-SA allows global-local feature interactions on various feature resolutions throughout the encoder-decoder architecture. In the proposed encoder-decoder architecture, both the window and grid sizes were fixed to \(8\) to make it compatible with \(256 \times 256\) image size. The number of attention heads was set to \(32\) for all attention blocks.

\subsection{MaxViT-UNet Encoder}\label{Encoder}

The encoder of the proposed MaxViT-UNet framework is made of MaxViT architecture \cite{tu2022maxvit} by simply stacking MBConv and Max-SA modules alternatively in a hierarchical fashion. Unlike the MaxViT \cite{tu2022maxvit}, where the number of blocks and channel dimensions are increased per stage to scale up the model. We used two MaxViT blocks per stage to obtain a small and efficient encoder. Additionally, the third stage was repeated 2 times and 5 times for the MoNuSeg18 and MoNuSAC20 datasets, respectively. The multi-class nature of the MoNuSAC20 dataset demands higher-level discriminating features obtained by repeating the third stage 5 times. The four stages of our encoder produce hierarchical feature representation just like UNet. MaxViT takes advantage of the local-global receptive fields via convolution and local-global attention mechanisms throughout the encoder from earlier to deeper stages and shows better generalization ability and model capacity. The last stage of the encoder is named bottleneck, as it contains semantic-rich features and provides a bridge from encoder to decoder.

\subsection{Proposed Multi-Axis Attention-based Hybrid Decoder Block}\label{Decoder}

\begin{figure}[ht!]
  \centering
  \includegraphics[width=\linewidth]{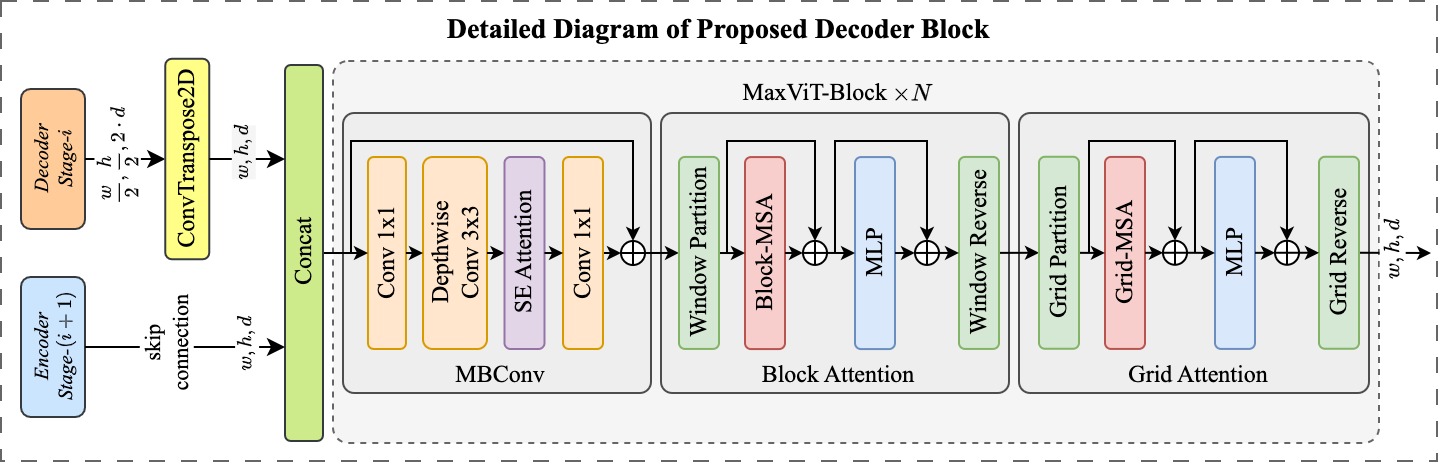}
  \caption{Detailed architecture of the proposed Hybrid Decoder block. Features from the  \(i^{th}\) decoder stage are upscaled using ConvTranspose2D layer to match with size of the \((i+1)^{th}\) encoder stage coming from skip-connection. After the concatenation (concat) operation, the MaxViT-block is used a couple of times to merge the features efficiently.}
  \label{fig:maxvit-unet-decoder-block}
\end{figure}

The proposed Hybrid Decoder is designed by stacking layers of Mutil-Axis Attention-based MaxViT-blocks in a hierarchical architecture, with a TransposeConv layer at the start of each stage, as shown in Figure \ref{fig:maxvit-unet-decoder-block}. Similar to the encoder, we created a parameter-efficient decoder by using only two MaxViT blocks per stage. The decoder also enjoys the global and local receptive fields at all stages and is able to better reconstruct output masks as compared to previous approaches. Similar to Swin-UNet \cite{cao2022swin}, our decoder contains three stages that are connected with the corresponding top three stages of the encoder. Features from the preceding decoder layer are transmitted through the TransposeConv layer inside a single decoder block in order to up-sample and match their shape with skip-path features. Semantically and spatially rich features are obtained by concatenating the up-sampled features with the associated skip-connection features. The MaxViT blocks further enhance them using MBConv, local attention, and global attention sub-block.

Let \(\textbf{y}^{\textbf{(i-1)}}\) represent the features coming from the previous decoder stage having dimension \(C \times H \times W\), and \(\textbf{z}^{\textbf{(i)}}\) represent features coming from skip-connection at the same stage having dimension \(C \times 2H \times 2W\), then the following equations represent the first block of each decoder stage:
\begin{align}
\textbf{y}^{\textbf{(i)}} &= \texttt{UPCONV}(\textbf{y}^{\textbf{(i-1)}}) \\
\textbf{y}^{\textbf{(i)}} &= \texttt{GRID}(\texttt{BLOCK}(\texttt{MBCONV}(\texttt{CONCAT}(\textbf{y}^{\textbf{(i)}}, \textbf{z}^{\textbf{(i)}}))))
\end{align}

where \texttt{UPCONV}, consists of TransposeConv layer, BatchNorm layer \cite{ioffe2015batch} and Mish activation function \cite{misra2019mish}, the \texttt{CONCAT} operator represents concatenation, and MBConv, \texttt{GRID}, and \texttt{BLOCK} are sub-blocks of MaxViT-block for feature processing.

\subsection{Loss Functions of the Proposed MaxViT-UNet}\label{Loss Functions of the Proposed MaxViT-UNet}

The models are penalized using the composite weighted loss function consisting of CrossEntropy Loss and Dice Loss functions with weights \(\lambda1\) and \(\lambda2\) respectively. In all the experiments, we used loss weights of \(\lambda1 = 1\) and \(\lambda2 = 3\). Both CrossEntropy and Dice losses are calculated pixel-wise. In the Dice loss, only the nuclei classes were considered and the background was ignored to force the model to focus on nuclei regions more strongly. Given true mask \(\textbf{y}_{t}\) and predicted mask \(\hat{\textbf{y}}_{p}\) images, the mathematical form of both the loss functions is as follows:
\begin{align}
\texttt{Loss}(\textbf{y}_{t}, \hat{\textbf{y}}_{p}) &= \lambda1 * \texttt{CELoss}(\textbf{y}_{t}, \hat{\textbf{y}}_{p}) + \lambda2 * \texttt{DiceLoss}(\textbf{y}_{t}, \hat{\textbf{y}}_{p}) \\
\texttt{CELoss}(\textbf{y}_{t}, \hat{\textbf{y}}_{p}) &= -\sum_{i=1}^{H \times W} (\textbf{y}_{t}^{i} * \texttt{log}(\hat{\textbf{y}}_{p}^{i})) \\
\texttt{DiceLoss}(\textbf{y}_{t}, \hat{\textbf{y}}_{p}) &= 1 - \sum_{c=1}^{C} 2 \times \frac{|\textbf{y}_{t}^{c} \cap \hat{\textbf{y}}_{p}^{c}|}{|\textbf{y}_{t}^{c}| + |\hat{\textbf{y}}_{p}^{c}|}
\end{align}

In \texttt{CELoss}, \(\textbf{y}_{t}^{i}\) represents the \(\textbf{i}^{\textbf{th}}\) pixel of the true mask image and \(\hat{\textbf{y}}_{p}^{i}\) represents the \(\textbf{i}^{\textbf{th}}\) pixel of the predicted mask image, and summation is performed over all pixels (\(H \times W\)) to accumulate the error for a complete image. In \texttt{DiceLoss}, \(\textbf{y}_{t}^{c}\) represents the \(\textbf{c}^{\textbf{th}}\) class channel of true mask image, and \(\hat{\textbf{y}}_{p}^{c}\) represents the \(\textbf{c}^{\textbf{th}}\) class channel of predicted mask image, and summation is performed over all classes (C) to accumulate the error for all classes and all pixels.

\section{Experiments and Results}\label{Experiments and Results}

Extensive experiments were conducted on medical image segmentation tasks to illustrate the effectiveness of the proposed Hybrid Decoder Network and the MaxViT-UNet segmentation framework. The dataset used, the pre-processing procedures carried out, the workspace, the hyper-parameters, and the performance measures used for assessment are all described in depth in the section that follows. Lastly, a comparison is made between the MaxViT-UNet's quantitative and qualitative outcomes with earlier image segmentation methods.

\subsection{Dataset Description}\label{Dataset Description}

To advance the research in this area, numerous competitions for medical image segmentation tasks have been organized during the last few years. We decided to use the MoNuSeg 2018 \cite{kumar2019multi} and MoNuSAC 2020 \cite{verma2021monusac2020} challenge datasets to demonstrate the efficacy of our suggested MaxViT-UNet system. The information for both datasets is summarized in Table \ref{table:dataset-summary}. Both datasets have their own challenges and deal with varying degrees of issues. The details of both datasets are highlighted in the following sections.

\begin{table}[ht!]
  \centering
  \caption{Summary of the datasets used to train and evaluate the proposed MaxViT-UNet.}\label{table:dataset-summary}
  \begin{tabular*}{\textwidth}{@{\extracolsep\fill}cccccc}
    \toprule%
      Dataset & Classes & Subset & Images & Nuclei & Organs \\
    \hline
      \multirow{4}{*}{MoNuSeg18} & \multirow{4}{*}{\begin{tabular}{@{\ }c@{}} Background, \\Nuclei \end{tabular}} & Train & 30 & 21,623 & \begin{tabular}{@{\ }c@{}} Breast, Kidney, Liver, \\Bladder, Colon, Stomach, \\Prostate \end{tabular} \\
      \cmidrule{3-6}
      & & Test & 14 & 7,223 & \begin{tabular}{@{\ }c@{}} Breast, Bladder, Kidney, \\Colon, Brain, Lung, \\Prostate \end{tabular} \\
    \hline
      \multirow{4}{*}{MoNuSAC20} & \multirow{4}{*}{\begin{tabular}{@{\ }c@{}} Epithelial, \\Lymphocytes, \\Macrophoges, \\Neutrophils \end{tabular}} & Train & 46 & 31,411 & \begin{tabular}{@{\ }c@{}} Breast, Kidney, \\Lung, Prostate \end{tabular} \\
      \cmidrule{3-6}
      & & Test & 25 & 15,498 & \begin{tabular}{@{\ }c@{}} Breast, Kidney, \\Lung, Prostate \end{tabular} \\
    \hline
  \end{tabular*}
\end{table}

\subsubsection{MoNuSeg18}

The MoNuSeg 2018 challenge provided a challenging dataset \cite{kumar2019multi} comprising images from 7 different organs: (1) breast, (2) colon, (3) bladder, (4) stomach, (5) kidney, (6) liver, and (7) prostate. Also, images acquired from 18 different hospitals, practicing different staining techniques and image acquisition equipment, add another source of variation and ensure the diversity of nuclear appearances. The training data consists of 30 tissue images (1000$\times$1000 resolution), 7 validation images, and 14 test images. The training dataset consists of 21623 annotated manually nuclear boundaries. For each selected individual patient from TCGA \cite{tcga}, an image was extracted from a distinct whole slide image (WSI) that was scanned at 40$\times$ magnification. Sub-images were selected from regions containing a high density of nuclei. To ensure diversity in the dataset, only one crop per WSI and patient was included. The test comprises 14 images spanning 5 organs common with the training set: (1) breast, (2) colon, (3) bladder, (4) kidney, (5) liver, and 2 organs different from the testing set: (1) lung, (2) brain, to make the test set more challenging. The test set contains approximately 7,223 annotated nuclear boundaries.

\subsubsection{MoNuSAC20}

The MoNuSAC20 dataset \cite{verma2021monusac2020} was designed to be representative of various organs and nucleus types relevant to tumor research. Specifically, it included Lymphocytes, Epithelial, Macrophages, and Neutrophils. The training data consisted of cropped whole slide images (WSIs) obtained from 32 hospitals and 46 patients from TCGA \cite{tcga} data portal, scanned at a 40$\times$ magnification. The dataset provides nuclei class labels along with nuclear boundary annotations. The testing data followed a similar preparation procedure but included annotations for ambiguous regions. These are regions with faint nuclei, unclear boundaries, or where the true class is not confirmed by annotators. The testing data comprised 25 patient samples from 19 different hospitals, with 14 hospitals overlapping with the training dataset.

\subsection{Dataset Pre-processing}\label{Dataset Pre-processing}

\subsubsection{Pre-processing of MoNuSeg18 Dataset}

\begin{figure}[ht!]
  \centering
  \includegraphics[width=\linewidth]{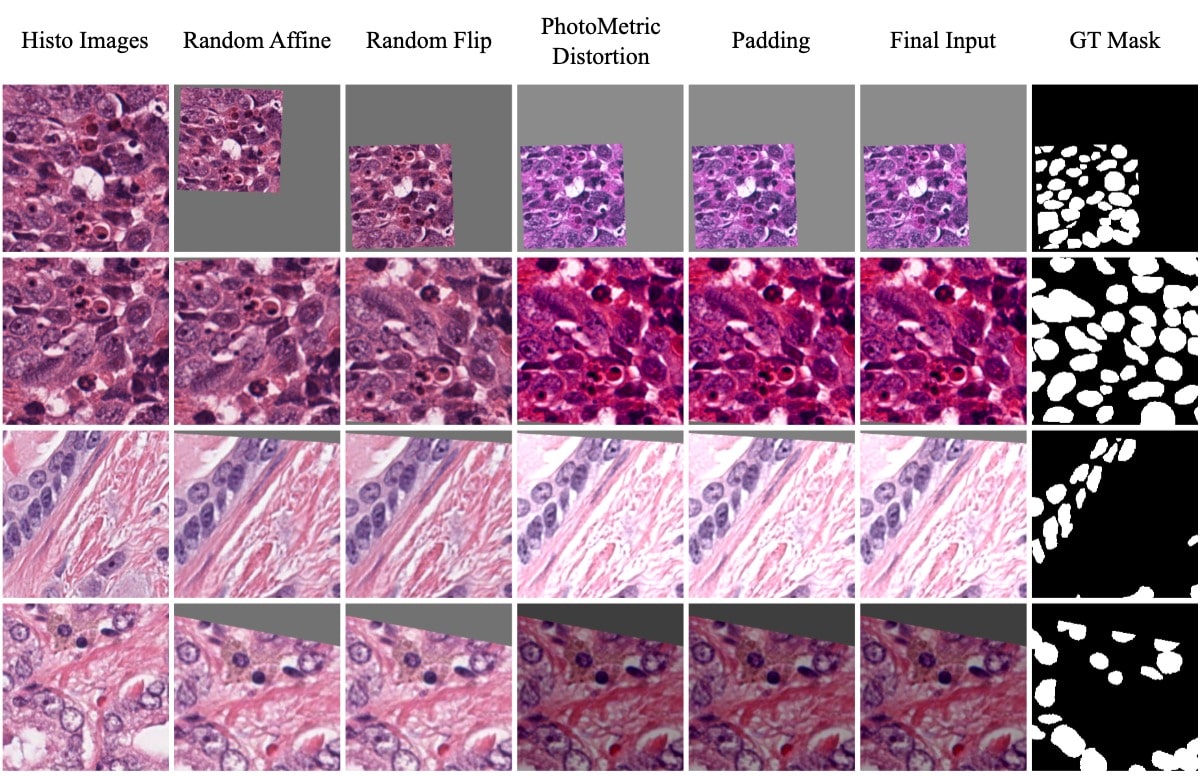}
  \caption{Data Pre-processing Pipeline visualized for MoNuSeg18 dataset. From left to right: Original Image (resized to \(256 \times 256\)), Random Affine (combination of Shift, Scale, and Rotate), Random Flip (either Horizontal or Vertical), PhotoMetric Distortion (changes the intensity of pixels), Padding (to ensure \(256 \times 256\) image size), Final Augmented Input and Mask image are shown.}
  \label{fig:data-pipeline}
  \vspace{-1.0em}
\end{figure}

For the MoNuSeg18 dataset \cite{kumar2019multi}, \(256 \times 256\) dimension patches (images and masks) from \(1000 \times 1000\) images were extracted to use for training and testing purposes of segmentation models. In order to prevent testing set leakage and inaccurate assessment metrics, it was also made sure that testing patches stayed in the testing set and training patches stayed in the training set. The size of the dataset is increased by employing a variety of augmentation techniques during training step, such as \textit{RandomAffine}, \textit{PhotoMetricDistortion}, \textit{Random Horizontal} and \textit{Vertical Flip} with \(0.5\) flip probability. The step-by-step outcome of these pre-processing steps is shown in figure \ref{fig:data-pipeline} for MoNuSeg18 \cite{kumar2019multi} dataset. Considering the modality differences between ImageNet and histopathology images, we calculated normalization parameters \texttt{(mean=[171.31, 119.69, 157.71], std=[56.04, 59.61, 47.69])} and used them for image normalization during training and testing phases.

\subsubsection{Pre-processing of MoNuSAC20 Dataset}

For the MoNuSAC20 dataset \cite{verma2021monusac2020}, the same pre-processing was applied as for the MoNuSeg18, i.e. \(256 \times 256\) dimension patches (images and masks) were extracted for training and testing. The same augmentation techniques were applied to increase the dataset size and robustness of the model. During the training and testing stages of MoNuSAC20, the ImageNet normalization parameters \texttt{mean=[123.675, 116.28, 103.53], std=[58.395, 57.12, 57.375]} were applied because they produced good results on this dataset.

\subsection{Working Environment}\label{Working Environment}

For the implementation, training, and evaluation of our proposed MaxViT-UNet and baseline models, we used MMSegmentation \cite{contributors2020openmmlab} (v0.24.1) and PyTorch \cite{pytorch} (v1.12.1) frameworks. We used conda (v4.12.0) for setting up our environment. All the trainings were done using NVIDIA DGX Station with 4 Tesla V100 GPUs, 120GB GPU memory, 256 GB RAM and Intel Xeon E5-2698 CPU.

\subsection{Training Details of the Proposed MaxViT-UNet}\label{Training Details of the Proposed MaxViT-UNet}

We made use of MMSegmentation's distributed training to achieve quick training speeds \cite{contributors2020openmmlab}. Due to distributed training across 4 GPUs, the batch size for a single GPU was set to 4, but the actual batch size was 16 instead. We conducted experiments using SGD, Adam \cite{kingma2014adam}, AdaBelief \cite{zhuang2020adabelief}, and AdamW \cite{loshchilov2018decoupled} optimizers. The outcomes of AdamW \cite{loshchilov2018decoupled} were superior to those of the other optimizers. In order to optimize our model through back-propagation, we utilized AdamW for all of our final training. We set the weight decay to 0.01 and the initial learning rate to 0.005. We also set the values of the betas to (0.9, 0.999). To gradually lower the learning rate and enable the model to stabilize at the optima, the Cosine learning rate scheduler was utilized.

\subsection{Performance Metrics}\label{Performance Metrics}

The MoNuSeg18 and MoNuSAC20 datasets were evaluated using Dice and IoU evaluation metrics. Both the Dice and IoU are widely used segmentation metrics that produce values between 0 and 1. The Dice is equivalent to F1-Score in image segmentation tasks, and IoU is also referred to as the Jaccard Index. Mathematically, Dice and IoU are defined as follows:
\begin{align}
\texttt{Dice}(\textbf{y}_{t}, \hat{\textbf{y}}_{p}) &= \sum_{c=1}^{C} 2 \times \frac{|\textbf{y}_{t}^{c} \cap \hat{\textbf{y}}_{p}^{c}|}{|\textbf{y}_{t}^{c}| + |\hat{\textbf{y}}_{p}^{c}|} \\
\texttt{IoU}(\textbf{y}_{t}, \hat{\textbf{y}}_{p}) &= \sum_{c=1}^{C} \frac{|\textbf{y}_{t}^{c} \cap \hat{\textbf{y}}_{p}^{c}|}{|\textbf{y}_{t}^{c} \cup \hat{\textbf{y}}_{p}^{c}|}
\end{align}

where \(\textbf{y}_{t}^{c}\) represents the \(\textbf{c}^{\textbf{th}}\) class channel of true mask image and \(\hat{\textbf{y}}_{p}^{c}\) represents the \(\textbf{c}^{\textbf{th}}\) class channel of predicted mask image. The summation is performed over all classes (C) to accumulate the evaluation metric for a complete image.

\subsection{Results and Discussions}\label{Results and Discussions}

The proposed MaxViT-UNet is compared with previous techniques on both the MoNuSeg18 and MoNuSAC20 datasets. The following sections discuss the experimental results on each dataset in detail.

\subsubsection{Performance Evaluation of the Proposed MaxViT-UNet}\label{Performance Evaluation of the Proposed MaxViT-UNet}

\begin{table}
  \centering
  \caption{Comparative results of the proposed MaxViT-UNet framework with previous techniques on MoNuSeg 2018 Challenge Dataset}\label{table:results-monuseg18}
  \begin{tabular*}{\textwidth}{@{\extracolsep\fill}cccccc}
    \toprule%
      Method & Dice & IoU  \\ 
    \hline
      U-Net \cite{ronneberger2015u} & 0.8185 & 0.6927 \\
      U-net++ \cite{zhou2019unet++} & 0.7528 & 0.6089 \\
      AttentionUnet \cite{oktay2018attention} & 0.7620 & 0.6264 \\
      MultiResUnet \cite{ibtehaz2020multiresunet} & 0.7754 & 0.6380 \\
      Bio-net \cite{xiang2020bio} & 0.7655 & 0.6252 \\
      TransUnet \cite{chen2021transunet} & 0.7920 & 0.6568 \\
      ATTransUNet \cite{li2023attransunet} & 0.7916 & 0.6551 \\
      MedT \cite{valanarasu2021medical} & 0.7924 & 0.6573 \\
      UCTransnet \cite{wang2022uctransnet} & 0.7987 & 0.6668 \\
      FSA-Net \cite{zhan2023fsa} & 0.8032 & 0.6699 \\
      MBUTransUNet \cite{qiao2023mbutransnet} & 0.8160 & 0.6902 \\
      DSREDN \cite{chanchal2022deep} & 0.8065 & - \\
      Swin-Unet \cite{cao2022swin} & 0.7956 & 0.6471 \\
      \textbf{MaxViT-UNet (Proposed)} & \textbf{0.8378} & \textbf{0.7208} \\
    \hline
  \end{tabular*}
\end{table}

\begin{table}
  \centering
  \caption{Comparative results of the proposed MaxViT-UNet framework with previous techniques on MoNuSAC 2020 Challenge Dataset}\label{table:results-monusac20}
  \begin{tabular*}{\textwidth}{@{\extracolsep\fill}cccccc}
    \toprule%
      Method & Dice & IoU  \\ 
    \hline
      UNet \cite{ronneberger2015u} & 0.7197 & 0.5874 \\
      Hover-net \cite{wang2023improved} & 0.7626 & - \\
      Dilated Hover-net w/o ASPP \cite{wang2023improved} & 0.7571 & - \\
      Dilated Hover-net w/ ASPP \cite{wang2023improved} & 0.7718 & - \\
      MulVerNet \cite{vo2023mulvernet} & 0.7660 & - \\
      NAS-SCAM \cite{liu2020scam} & 0.6501 & - \\
      PSPNet \cite{zhao2017pyramid} & 0.7893 & 0.6594 \\
      Swin-Unet \cite{cao2022swin} & 0.4689 & 0.3924 \\
      \textbf{MaxViT-UNet (Proposed)} & \textbf{0.8215} & \textbf{0.7030} \\
    \hline
  \end{tabular*}
\end{table}

\begin{figure}[ht!]
  \centering
  \begin{subfigure}[]{0.48\textwidth}
    \centering
    \includegraphics[width=\textwidth]{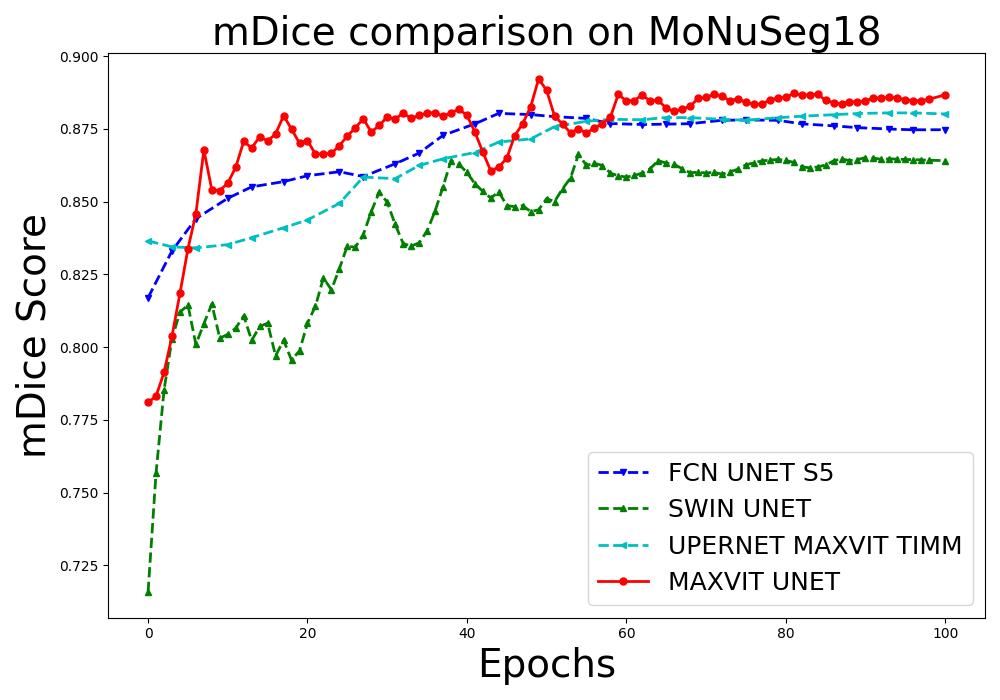}
    \caption{}
    \label{fig:monuseg18-mDice}
  \end{subfigure}
  \hfill
  \begin{subfigure}[]{0.48\textwidth}
    \centering
    \includegraphics[width=\textwidth]{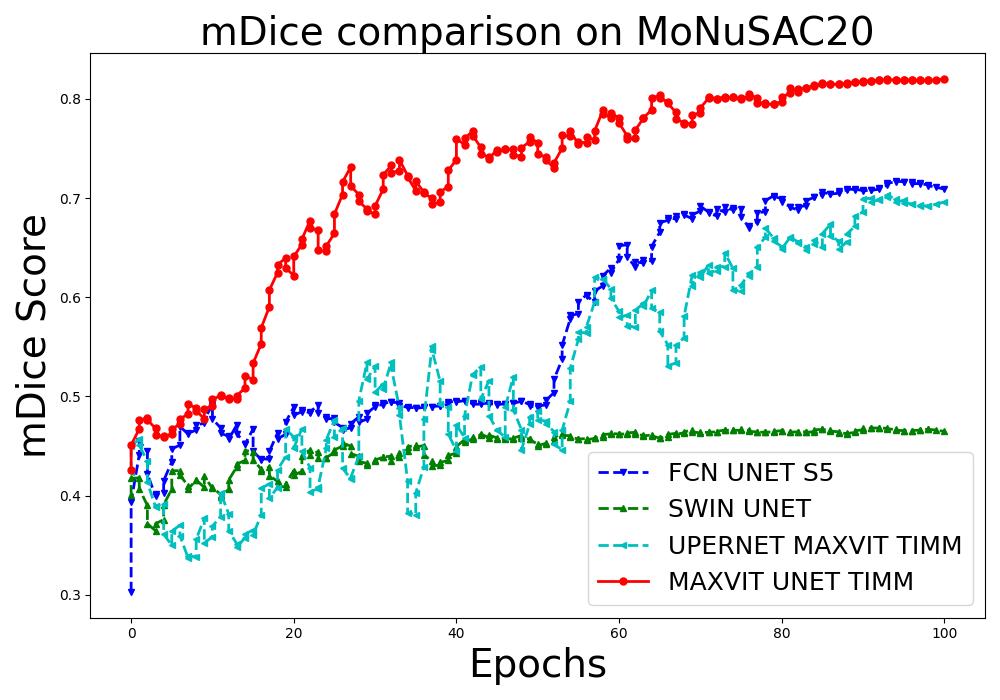}
    \caption{}
    \label{fig:monusac20-mDice}
  \end{subfigure}
  \hfill
  \begin{subfigure}[]{0.48\textwidth}
    \centering
    \includegraphics[width=\textwidth]{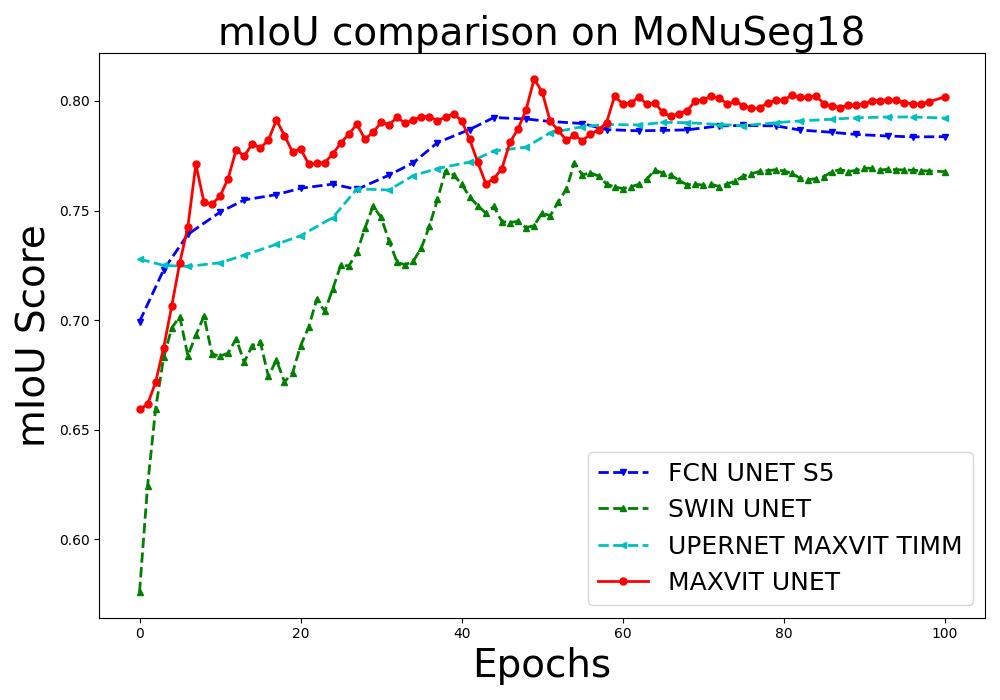}
    \caption{}
    \label{fig:monuseg18-mIoU}
  \end{subfigure}
  \hfill
  \begin{subfigure}[]{0.48\textwidth}
    \centering
    \includegraphics[width=\textwidth]{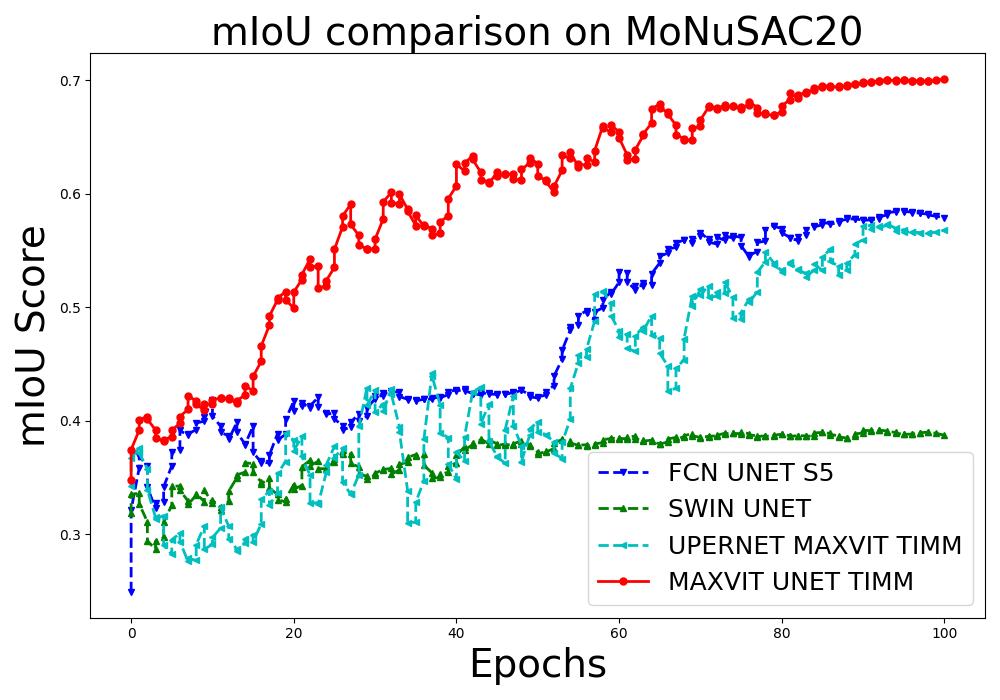}
    \caption{}
    \label{fig:monusac20-mIoU}
  \end{subfigure}
  \hfill
  \begin{subfigure}[]{0.48\textwidth}
    \centering
    \includegraphics[width=\textwidth]{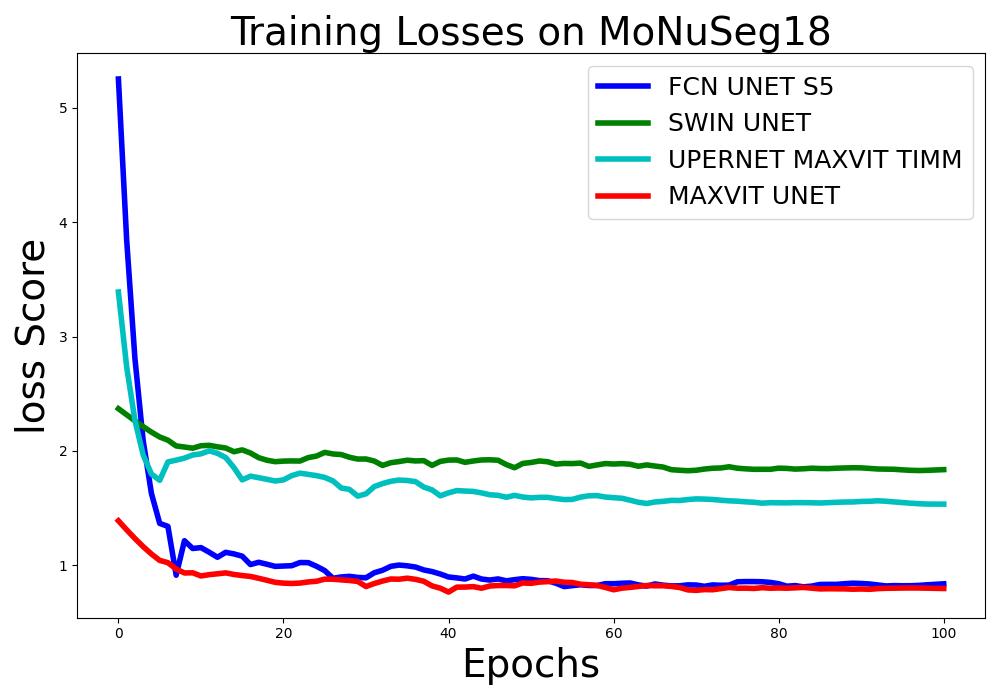}
    \caption{}
    \label{fig:monuseg18-loss}
  \end{subfigure}
  \hfill
  \begin{subfigure}[]{0.48\textwidth}
    \centering
    \includegraphics[width=\textwidth]{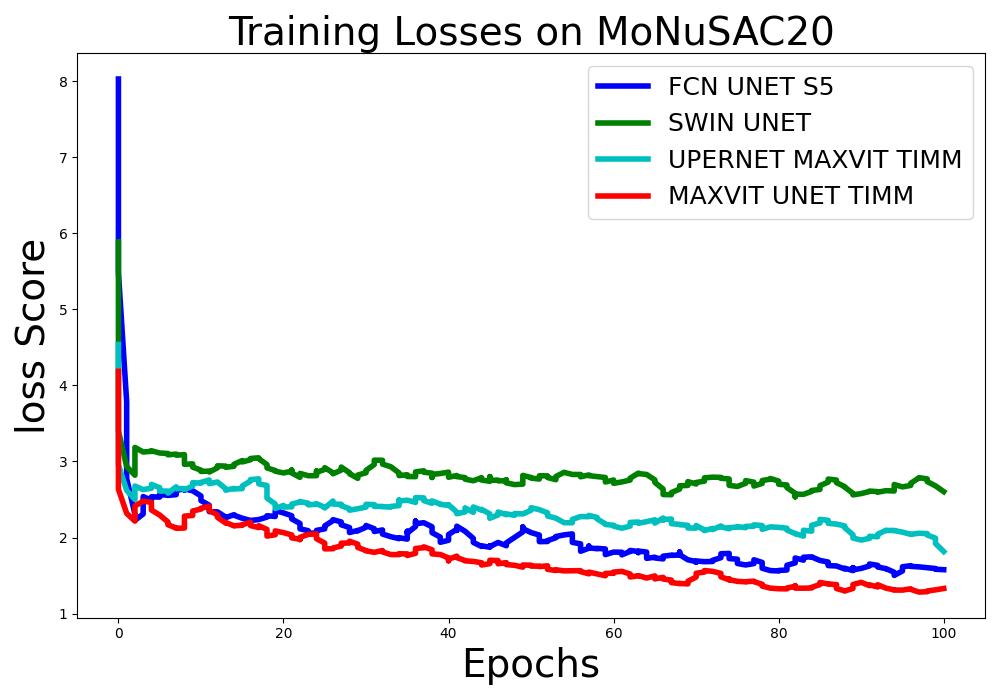}
    \caption{}
    \label{fig:monusac20-loss}
  \end{subfigure}
  \caption{Comparative plots of the proposed MaxViT-UNet with previous techniques on MoNuSeg18 and MoNuSAC20 challenge datasets. The left column displays the (a) Dice, (b) IoU, and (c) Training Loss on the MoNuSeg18 dataset, whereas the right column displays the (d) Dice, (e) IoU and (f) Training Loss on MoNuSAC20 dataset.}
  \label{fig:metric-plots}
  \vspace{-1.0em}
\end{figure}

\begin{figure}[ht!]
  \centering
  \begin{subfigure}[]{0.48\textwidth}
    \centering
    \includegraphics[width=\textwidth]{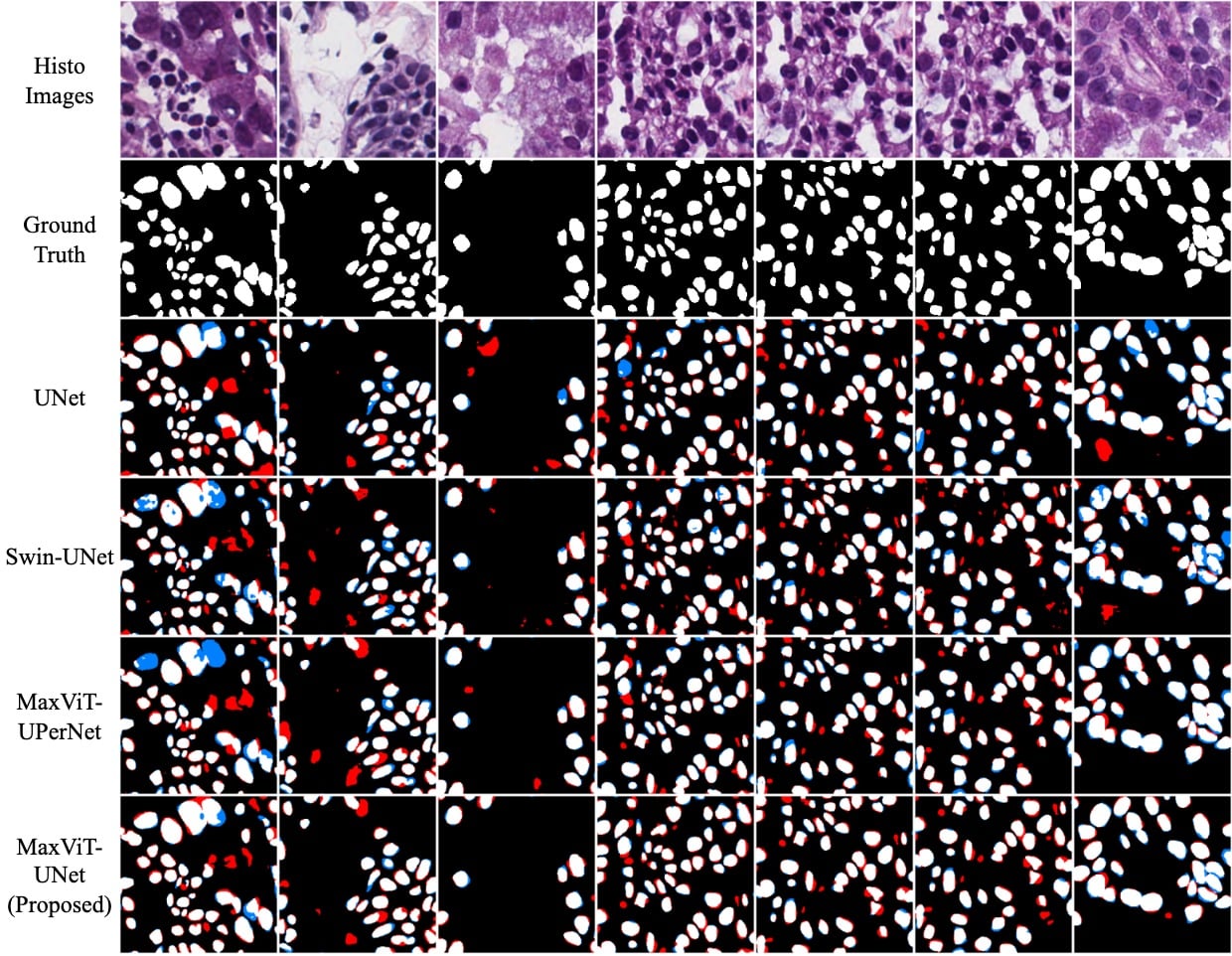}
    \caption{}
    \label{fig:qualitative-monuseg-01}
  \end{subfigure}
  \hfill
  \begin{subfigure}[]{0.48\textwidth}
    \centering
    \includegraphics[width=\textwidth]{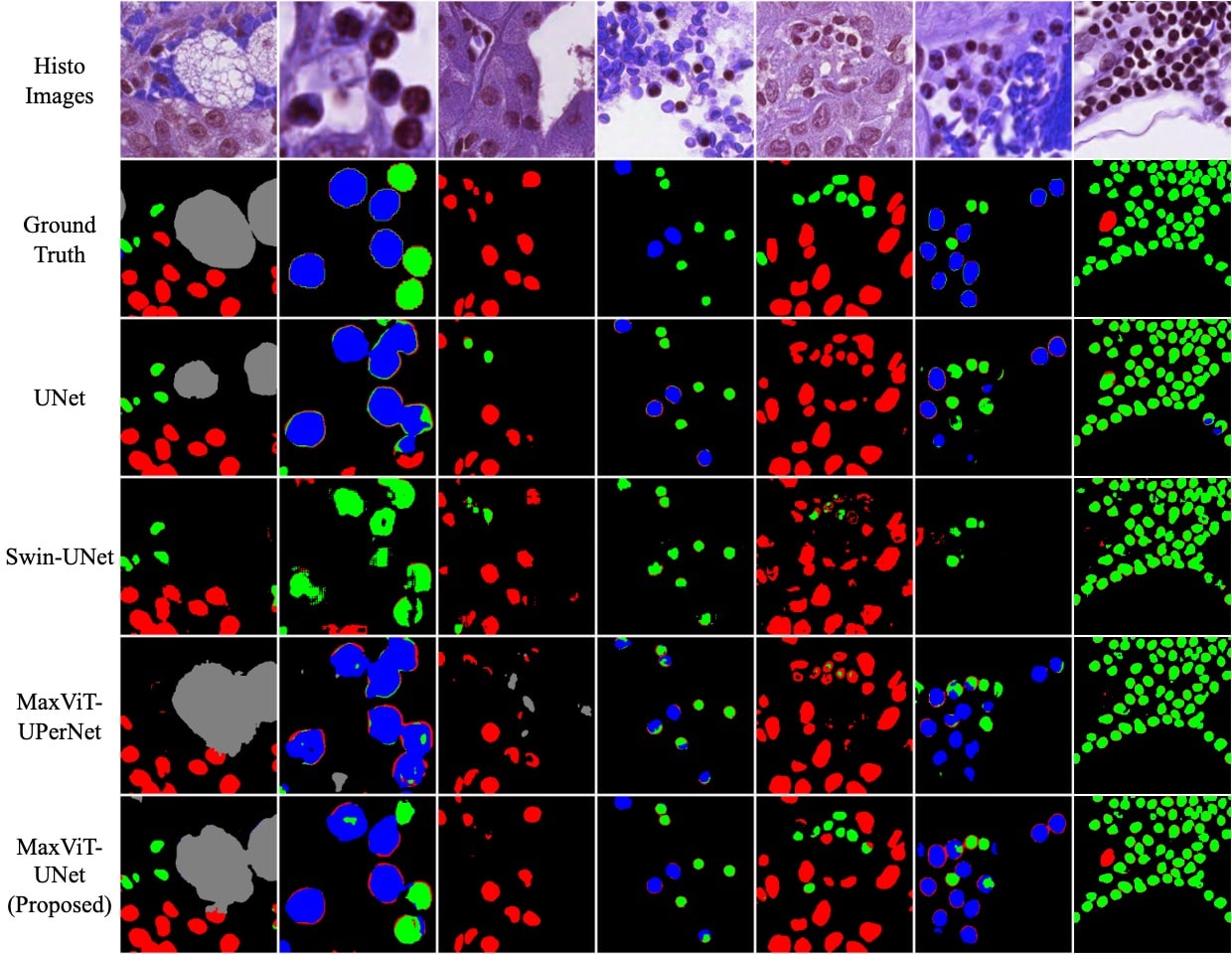}
    \caption{}
    \label{fig:qualitative-monusac-01}
  \end{subfigure}
  \caption{Qualitative comparison of the proposed MaxViT-UNet with current methods on (a) MoNuSeg18 dataset; the colors white, red, and blue, respectively, indicate True-Positive, False-Positive, and False-Negative predictions. (b) The MoNuSAC20 dataset shows red, yellow, green, and blue representations of epithelial, lymphocyte, macrophage, and neutrophil, respectively.}
  \label{fig:qualitative-comparision}
\end{figure}

\begin{figure}[ht!]
  \centering
  \begin{subfigure}[]{0.48\textwidth}
    \centering
    \includegraphics[width=\textwidth]{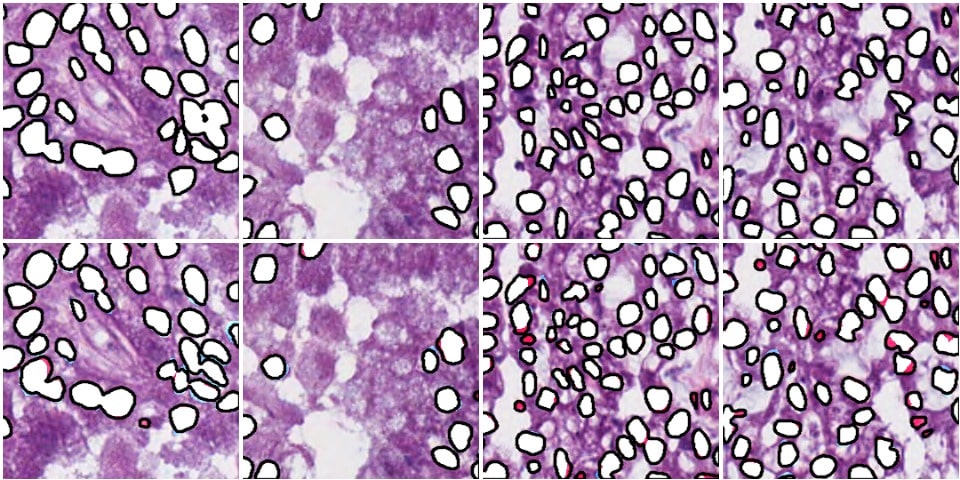}
    \caption{}
    \label{fig:qualitative-monuseg-02}
  \end{subfigure}
  \hfill
  \begin{subfigure}[]{0.48\textwidth}
    \centering
    \includegraphics[width=\textwidth]{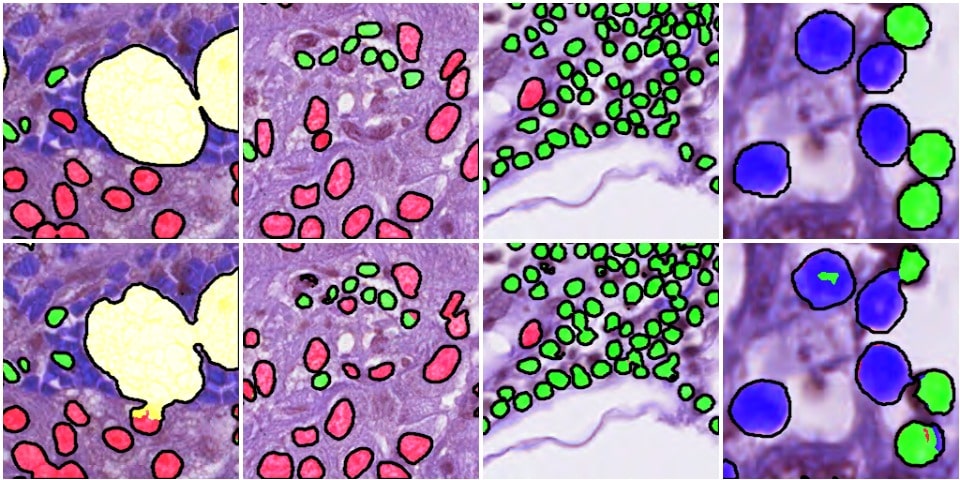}
    \caption{}
    \label{fig:qualitative-monusac-02}
  \end{subfigure}
  \caption{Comparison of the MaxViT-UNet predicted masks (bottom rows) and the True masks (top rows) overlaid over the histopathological images on (a) the MoNuSeg18 dataset; the colors red, blue, and white, respectively, indicate false-negative, false-positive, and true-positive predictions. (a) MoNuSAC20 dataset; red, yellow, green, and blue denote epithelial, lymphocyte, macrophage, and neutrophil, respectively.}
  \label{fig:qualitative-overlay}
  \vspace{-1.0em}
\end{figure}

The comparison between the proposed MaxViT-UNet and previous methodologies is presented in Table \ref{table:results-monuseg18} on the MoNuSeg18 dataset and Table \ref{table:results-monusac20} on the MoNuSAC20 dataset. For comparison on both datasets, UNet and Swin-UNet were trained using MMSegmentation \cite{contributors2020openmmlab} with the same hyper-parameters as the proposed technique. For the MoNuSeg18 dataset, we performed binary semantic segmentation. Whereas the MoNuSAC20 challenge contains four types of nuclei, we performed multi-class semantic segmentation for the MoNuSAC20 dataset. The proposed MaxViT-UNet beats the previous techniques by a large margin on both datasets and proves the significance of the hybrid encoder-decoder architecture.

On MoNuSeg18 dataset, the CNN-based UNet achieved 0.8185 Dice and 0.6927 IoU scores. Whereas, the Transformer-based Swin-Unet achieved 0.7956 Dice and 0.6471 IoU scores. In comparison, our proposed hybrid framework MaxViT-UNet is able to achieve superior scores for both Dice (0.8378) and IoU (0.7208) metrics. It surpassed the CNN-based UNet \cite{ronneberger2015u} by 2.36\% Dice score and 4.06\% IoU score; and Transformer-based Swin-UNet \cite{cao2022swin} by 5.31\% Dice score and 11.40\% IoU score on MoNuSeg18 dataset.

For the MoNuSAC20 dataset, the CNN-based UNet achieved 0.7197 mDice (mean Dice) and 0.5874 mIoU (mean IoU) scores. Whereas, the Transformer-based Swin-Unet achieved 0.4689 Dice and 0.3924 IoU scores. In comparison, our proposed hybrid framework MaxViT-UNet is able to achieve superior scores for both Dice (0.8215) and IoU (0.7030) metrics. It surpassed CNN-based UNet \cite{ronneberger2015u} by 14.14\% Dice and 19.68\% IoU scores; and Transformer-based Swin-UNet \cite{cao2022swin} by a large margin on Dice and IoU metrics as evident from Table \ref{table:results-monusac20}. The large improvement in both mDice and mIoU scores shows the significance of hybrid encoder-decoder architecture.


The learning curve plots of Dice, IoU, and training loss on MoNuSeg18 dataset are shown in figs. \ref{fig:monuseg18-mDice}, \ref{fig:monuseg18-mIoU}, \ref{fig:monuseg18-loss}, whereas mDice, mIoU, and training loss on MoNuSAC20 dataset are shown in figs. \ref{fig:monusac20-mDice}, \ref{fig:monusac20-mIoU}, \ref{fig:monusac20-loss} respectively. The proposed MaxViT-UNet framework is represented by red curve lines in all the mentioned figures. The baselines are shown with different colors that are the same throughout all the metric plots, e.g. the UNet is represented with blue curve lines and Swin-UNet is represented with green curve lines. The third baseline, MaxViT with UPerNet decoder, is represented with light blue curve lines and it's details are discussed in ablation study section below. The plots for Dice and IoU on both MoNuSeg18 and MoNuSAC20 dataset shows that the proposed MaxViT-UNet was able to obtain optimal performance in the initial training epochs and maintained its superiority over baselines models. This swift performance can be attributed to the hybrid nature of the proposed framework and its ability to capture local and global features simultaneously. In comparison, the Swin-UNet showed poor performance among the baselines in all plots, maybe due to the fact that it relies totally on the self-attention mechanism and lacks the inductive bias of convolution operation. The training loss curve of the proposed MaxViT-UNet is also very stable and lower than the baselines in both datasets, showing the stability and convergence of the proposed framework.

The qualitative results on diverse images are presented on the MoNuSeg18 dataset in fig. \ref{fig:qualitative-monuseg-01} and the MoNuSAC20 dataset in fig. \ref{fig:qualitative-monusac-01}. The masks generated by MaxViT-UNet are less prone to error and produce relatively accurate boundaries as compared to vanilla UNet \cite{ronneberger2015u} and Swin-UNet \cite{cao2022swin}. In fig. \ref{fig:qualitative-monuseg-01} for MoNuSeg18 dataset, the white color represents the true predicted regions, whereas red and blue colors highlight the erroneous regions. In fig. \ref{fig:qualitative-monusac-01} for the MoNuSAC20 dataset, four different colors represent four types of nuclei classes in the dataset: red represents epithelial, yellow corresponds to lymphocyte, green shows macrophage, and blue indicates neutrophil.

The figs. \ref{fig:qualitative-monuseg-02} and \ref{fig:qualitative-monusac-02} compare the ground truth mask images and predicted mask images of the proposed MaxViT-UNet overlaid on histopathology images. The color coding used in these figures to represent different nuclei regions is the same as that in figs. \ref{fig:qualitative-monuseg-01} and \ref{fig:qualitative-monusac-01}.

\subsubsection{Ablation Study of the Proposed MaxViT-UNet}\label{Ablation Study of the Proposed MaxViT-UNet}

A comparison analysis was carried out utilizing the MaxViT encoder in conjunction with the UPerHead Decoder \cite{xiao2018unified} network in order to assess the effectiveness of the suggested Hybrid Decoder. In the bottleneck, this decoder makes use of a Pyramid Pooling Module (PPM) \cite{zhao2017pyramid}, and it uses only convolutional layers for decoding. The Dice and IoU measurements obtained on the MoNuSeg18 and MoNuSAC20 datasets are shown in Table \ref{table:ablation-decoder}. These outcomes clearly show how successful the suggested hybrid decoder is. Interestingly, MoNuSAC20's multi-class problem showed a notable margin of improvement, indicating its greater capacity to handle and discriminate between areas belonging to different classes. The decoder's hybrid design, which allows it to effectively use both local and global contextual information at many scales, is probably the cause of this improved performance.

Furthermore, the symmetric design of the Hybrid Decoder allows for its standalone integration into various UNet-like encoder-decoder architectures. The achieved performance enhancements suggest its potential to generate accurate segmentation masks even when paired with different types of encoders. This versatility highlights its potential broader applicability across diverse medical imaging tasks.

\begin{table}[ht!]
  \centering
  \caption{Ablation study results of the proposed MaxViT-UNet Decoder}\label{table:ablation-decoder}
  \begin{tabular*}{\textwidth}{@{\extracolsep\fill}ccccc}
    \toprule%
      Method & Dataset & Image Size & Dice & IoU \\
    \midrule
      MaxViT with UPerNet Decoder & MoNuSeg18 & (256, 256) & 0.8176 & 0.6914 \\
    \midrule
      MaxViT-UNet (Proposed) & MoNuSeg18 & (256, 256) & \textbf{0.8378} & \textbf{0.7208} \\
    \midrule
      MaxViT with UPerNet Decoder & MoNuSAC20 & (256, 256) & 0.7148 & 0.5828 \\
    \midrule
      MaxViT-UNet (Proposed) & MoNuSAC20 & (256, 256) & \textbf{0.8215} & \textbf{0.7030} \\
    \hline
  \end{tabular*}
\end{table}

\section{Conclusion}\label{Conclusion}

This work proposes MaxViT-UNet, a novel hybrid encoder-decoder architecture based on the UNet framework for medical image segmentation. To complement the hybrid nature of the MaxViT-based encoder, we also proposed a novel Hybrid Architecture as a Decoder. The proposed Hybrid Decoder effectively utilizes the MaxViT-block, consisting of an MBConv convolution block followed by an efficient multi-axis attention mechanism (Max-SA), to generate accurate segmentation masks. The hybrid block approach in both the encoder and decoder stages enables end-to-end capturing of rich hierarchical features with local and global information at multiple scales. The proposed network, and especially the novel hybrid decoder, is lightweight, computationally efficient, and designed as a modular plug-and-play component for UNet-like architectures.

Tests conducted on the MoNuSeg18 and MoNuSAC20 datasets show how successful the new Hybrid Decoder architecture and suggested MaxViT-UNet framework are. In terms of Dice and IoU metrics, our method significantly beat earlier CNN-based (UNet) and Transformer-based (Swin-UNet) approaches on both datasets.

Future work will focus on extending the proposed framework and Hybrid Decoder to other 2D/3D imaging modalities and real-world datasets. Investigating techniques such as channel boosting and ensemble learning could further enhance the robustness and generalizability of the segmentation approach.

\section*{Acknowledgements}

We thank Pattern Recognition Lab (PR-Lab) and the Pakistan Institute of Engineering and Applied Sciences (PIEAS) for providing the necessary computational resources and a healthy research environment.

\section*{Declarations}

\subsection*{Funding/Competing interests}
The authors declare that they have no known competing financial interests or personal relationships that could have appeared to influence the work reported in this paper.

\subsection*{Availability of data and materials}
The datasets used in this work are publicly available.

\subsection*{Code availability}
The code is available on \href{https://github.com/PRLAB21/MaxViT-UNet}{github} (https://github.com/PRLAB21/MaxViT-UNet).

\bibliographystyle{unsrt}  
\bibliography{references}

\begin{thebibliography}{10}

\bibitem{kumar2019multi}
Neeraj Kumar, Ruchika Verma, Deepak Anand, Yanning Zhou, Omer~Fahri Onder,
  Efstratios Tsougenis, Hao Chen, Pheng-Ann Heng, Jiahui Li, Zhiqiang Hu,
  et~al.
\newblock A multi-organ nucleus segmentation challenge.
\newblock {\em IEEE transactions on medical imaging}, 39(5):1380--1391, 2019.

\bibitem{tayyab2022survey}
Umm-e-Hani Tayyab, Faiza~Babar Khan, Muhammad~Hanif Durad, Asifullah Khan, and
  Yeon~Soo Lee.
\newblock A survey of the recent trends in deep learning based malware
  detection.
\newblock {\em Journal of Cybersecurity and Privacy}, 2(4):800--829, 2022.

\bibitem{sohail2023deep}
Anabia Sohail, Bibi Ayisha, Irfan Hameed, Muhammad~Mohsin Zafar, and Asifullah
  Khan.
\newblock Deep neural networks based meta-learning for network intrusion
  detection.
\newblock {\em arXiv preprint arXiv:2302.09394}, 2023.

\bibitem{khan2023survey_covid}
Asifullah Khan, Saddam~Hussain Khan, Mahrukh Saif, Asiya Batool, Anabia Sohail,
  and Muhammad Waleed~Khan.
\newblock A survey of deep learning techniques for the analysis of covid-19 and
  their usability for detecting omicron.
\newblock {\em Journal of Experimental \& Theoretical Artificial Intelligence},
  pages 1--43, 2023.

\bibitem{Rauf2023-tc}
Zunaira Rauf, Abdul~Rehman Khan, Anabia Sohail, Hani Alquhayz, Jeonghwan Gwak,
  and Asifullah Khan.
\newblock Lymphocyte detection for cancer analysis using a novel fusion block
  based channel boosted cnn.
\newblock {\em Scientific Reports}, 13(1):14047, aug 2023.

\bibitem{khan2008machine}
Asifullah Khan, Syed~Fahad Tahir, Abdul Majid, and Tae-Sun Choi.
\newblock Machine learning based adaptive watermark decoding in view of
  anticipated attack.
\newblock {\em Pattern Recognition}, 41(8):2594--2610, 2008.

\bibitem{naveed2012gpcr}
Muhammad Naveed and Asif~Ullah Khan.
\newblock Gpcr-mpredictor: multi-level prediction of g protein-coupled
  receptors using genetic ensemble.
\newblock {\em Amino Acids}, 42:1809--1823, 2012.

\bibitem{khan2017random}
Saranjam Khan, Rahat Ullah, Asifullah Khan, Anabia Sohail, Noorul Wahab,
  Muhammad Bilal, and Mushtaq Ahmed.
\newblock Random forest-based evaluation of raman spectroscopy for dengue fever
  analysis.
\newblock {\em Applied spectroscopy}, 71(9):2111--2117, 2017.

\bibitem{chouhan2021deep}
Naveed Chouhan, Asifullah Khan, Jehan~Zeb Shah, Mazhar Hussnain, and
  Muhammad~Waleed Khan.
\newblock Deep convolutional neural network and emotional learning based breast
  cancer detection using digital mammography.
\newblock {\em Computers in Biology and Medicine}, 132:104318, 2021.

\bibitem{majid2006combination}
Abdul Majid, Asifullah Khan, and Anwar~M Mirza.
\newblock Combination of support vector machines using genetic programming.
\newblock {\em International Journal of Hybrid Intelligent Systems},
  3(2):109--125, 2006.

\bibitem{zahoor2022new}
Mirza~Mumtaz Zahoor, Shahzad~Ahmad Qureshi, Sameena Bibi, Saddam~Hussain Khan,
  Asifullah Khan, Usman Ghafoor, and Muhammad~Raheel Bhutta.
\newblock A new deep hybrid boosted and ensemble learning-based brain tumor
  analysis using mri.
\newblock {\em Sensors}, 22(7):2726, 2022.

\bibitem{long2015fully}
Jonathan Long, Evan Shelhamer, and Trevor Darrell.
\newblock Fully convolutional networks for semantic segmentation.
\newblock In {\em Proceedings of the IEEE conference on computer vision and
  pattern recognition}, pages 3431--3440, 2015.

\bibitem{ronneberger2015u}
Olaf Ronneberger, Philipp Fischer, and Thomas Brox.
\newblock U-net: Convolutional networks for biomedical image segmentation.
\newblock In {\em Medical Image Computing and Computer-Assisted
  Intervention--MICCAI 2015: 18th International Conference, Munich, Germany,
  October 5-9, 2015, Proceedings, Part III 18}, pages 234--241. Springer, 2015.

\bibitem{cao2022swin}
Hu~Cao, Yueyue Wang, Joy Chen, Dongsheng Jiang, Xiaopeng Zhang, Qi~Tian, and
  Manning Wang.
\newblock Swin-unet: Unet-like pure transformer for medical image segmentation.
\newblock In {\em European Conference on Computer Vision}, pages 205--218.
  Springer, 2022.

\bibitem{rauf2023attention}
Zunaira Rauf, Anabia Sohail, Saddam~Hussain Khan, Asifullah Khan, Jeonghwan
  Gwak, and Muhammad Maqbool.
\newblock Attention-guided multi-scale deep object detection framework for
  lymphocyte analysis in ihc histological images.
\newblock {\em Microscopy}, 72(1):27--42, 2023.

\bibitem{ali2022channel}
Momina~Liaqat Ali, Zunaira Rauf, Abdur~Rehman Khan, and Asifullah Khan.
\newblock Channel boosting based detection and segmentation for cancer analysis
  in histopathological images.
\newblock In {\em 2022 19th International Bhurban Conference on Applied
  Sciences and Technology (IBCAST)}, pages 1--6. IEEE, 2022.

\bibitem{aziz2020channel}
Abdullah Aziz, Anabia Sohail, Labiba Fahad, Muhammad Burhan, Noorul Wahab, and
  Asifullah Khan.
\newblock Channel boosted convolutional neural network for classification of
  mitotic nuclei using histopathological images.
\newblock In {\em 2020 17th International Bhurban Conference on Applied
  Sciences and Technology (IBCAST)}, pages 277--284. IEEE, 2020.

\bibitem{sohail2021mitotic}
Anabia Sohail, Asifullah Khan, Humaira Nisar, Sobia Tabassum, and Aneela
  Zameer.
\newblock Mitotic nuclei analysis in breast cancer histopathology images using
  deep ensemble classifier.
\newblock {\em Medical image analysis}, 72:102121, 2021.

\bibitem{khan2023segmentation}
Saddam~Hussain Khan, Asifullah Khan, Yeon~Soo Lee, Mehdi Hassan, and Woong~Kyo
  Jeong.
\newblock Segmentation of shoulder muscle mri using a new region and edge based
  deep auto-encoder.
\newblock {\em Multimedia Tools and Applications}, 82(10):14963--14984, 2023.

\bibitem{khan2023survey}
Asifullah Khan, Zunaira Rauf, Anabia Sohail, Abdul~Rehman Khan, Hifsa Asif,
  Aqsa Asif, and Umair Farooq.
\newblock A survey of the vision transformers and their cnn-transformer based
  variants.
\newblock {\em Artificial Intelligence Review}, 56(Suppl 3):2917--2970, 2023.

\bibitem{zhang2022multi}
Baihua Zhang, Shouliang Qi, Yanan Wu, Xiaohuan Pan, Yudong Yao, Wei Qian, and
  Yubao Guan.
\newblock Multi-scale segmentation squeeze-and-excitation unet with conditional
  random field for segmenting lung tumor from ct images.
\newblock {\em Computer Methods and Programs in Biomedicine}, 222:106946, 2022.

\bibitem{vu2019dense}
Quoc~Dang Vu and Jin~Tae Kwak.
\newblock A dense multi-path decoder for tissue segmentation in histopathology
  images.
\newblock {\em Computer methods and programs in biomedicine}, 173:119--129,
  2019.

\bibitem{liu2023mestrans}
Yatong Liu, Yu~Zhu, Ying Xin, Yanan Zhang, Dawei Yang, and Tao Xu.
\newblock Mestrans: Multi-scale embedding spatial transformer for medical image
  segmentation.
\newblock {\em Computer Methods and Programs in Biomedicine}, 233:107493, 2023.

\bibitem{khan2023recent}
Asifullah Khan, Zunaira Rauf, Abdul~Rehman Khan, Saima Rathore, Saddam~Hussain
  Khan, Najmus~Saher Shah, Umair Farooq, Hifsa Asif, Aqsa Asif, Umme Zahoora,
  et~al.
\newblock A recent survey of vision transformers for medical image
  segmentation.
\newblock {\em arXiv preprint arXiv:2312.00634}, 2023.

\bibitem{khan2020survey}
Asifullah Khan, Anabia Sohail, Umme Zahoora, and Aqsa~Saeed Qureshi.
\newblock A survey of the recent architectures of deep convolutional neural
  networks.
\newblock {\em Artificial intelligence review}, 53:5455--5516, 2020.

\bibitem{ibtehaz2020multiresunet}
Nabil Ibtehaz and M~Sohel Rahman.
\newblock Multiresunet: Rethinking the u-net architecture for multimodal
  biomedical image segmentation.
\newblock {\em Neural networks}, 121:74--87, 2020.

\bibitem{zhou2019unet++}
Zongwei Zhou, Md~Mahfuzur~Rahman Siddiquee, Nima Tajbakhsh, and Jianming Liang.
\newblock Unet++: Redesigning skip connections to exploit multiscale features
  in image segmentation.
\newblock {\em IEEE transactions on medical imaging}, 39(6):1856--1867, 2019.

\bibitem{oktay2018attention}
Ozan Oktay, Jo~Schlemper, Loic~Le Folgoc, Matthew Lee, Mattias Heinrich,
  Kazunari Misawa, Kensaku Mori, Steven McDonagh, Nils~Y Hammerla, Bernhard
  Kainz, Ben Glocker, and Daniel Rueckert.
\newblock Attention u-net: Learning where to look for the pancreas.
\newblock In {\em Medical Imaging with Deep Learning}, 2018.

\bibitem{cciccek20163d}
{\"O}zg{\"u}n {\c{C}}i{\c{c}}ek, Ahmed Abdulkadir, Soeren~S Lienkamp, Thomas
  Brox, and Olaf Ronneberger.
\newblock 3d u-net: learning dense volumetric segmentation from sparse
  annotation.
\newblock In {\em Medical Image Computing and Computer-Assisted
  Intervention--MICCAI 2016: 19th International Conference, Athens, Greece,
  October 17-21, 2016, Proceedings, Part II 19}, pages 424--432. Springer,
  2016.

\bibitem{dosovitskiy2021an}
Alexey Dosovitskiy, Lucas Beyer, Alexander Kolesnikov, Dirk Weissenborn,
  Xiaohua Zhai, Thomas Unterthiner, Mostafa Dehghani, Matthias Minderer, Georg
  Heigold, Sylvain Gelly, Jakob Uszkoreit, and Neil Houlsby.
\newblock An image is worth 16x16 words: Transformers for image recognition at
  scale.
\newblock In {\em International Conference on Learning Representations}, 2021.

\bibitem{chen2021transunet}
Jieneng Chen, Yongyi Lu, Qihang Yu, Xiangde Luo, Ehsan Adeli, Yan Wang, Le~Lu,
  Alan~L Yuille, and Yuyin Zhou.
\newblock Transunet: Transformers make strong encoders for medical image
  segmentation.
\newblock {\em arXiv preprint arXiv:2102.04306}, 2021.

\bibitem{valanarasu2021medical}
Jeya Maria~Jose Valanarasu, Poojan Oza, Ilker Hacihaliloglu, and Vishal~M
  Patel.
\newblock Medical transformer: Gated axial-attention for medical image
  segmentation.
\newblock In {\em Medical Image Computing and Computer Assisted
  Intervention--MICCAI 2021: 24th International Conference, Strasbourg, France,
  September 27--October 1, 2021, Proceedings, Part I 24}, pages 36--46.
  Springer, 2021.

\bibitem{wang2022uctransnet}
Haonan Wang, Peng Cao, Jiaqi Wang, and Osmar~R Zaiane.
\newblock Uctransnet: rethinking the skip connections in u-net from a
  channel-wise perspective with transformer.
\newblock In {\em Proceedings of the AAAI conference on artificial
  intelligence}, volume~36, pages 2441--2449, 2022.

\bibitem{zhang2021transfuse}
Yundong Zhang, Huiye Liu, and Qiang Hu.
\newblock Transfuse: Fusing transformers and cnns for medical image
  segmentation.
\newblock In {\em Medical Image Computing and Computer Assisted
  Intervention--MICCAI 2021: 24th International Conference, Strasbourg, France,
  September 27--October 1, 2021, Proceedings, Part I 24}, pages 14--24.
  Springer, 2021.

\bibitem{li2023attransunet}
Xuewei Li, Shuo Pang, Ruixuan Zhang, Jialin Zhu, Xuzhou Fu, Yuan Tian, and Jie
  Gao.
\newblock Attransunet: An enhanced hybrid transformer architecture for
  ultrasound and histopathology image segmentation.
\newblock {\em Computers in Biology and Medicine}, 152:106365, 2023.

\bibitem{guo2022cmt}
Jianyuan Guo, Kai Han, Han Wu, Yehui Tang, Xinghao Chen, Yunhe Wang, and Chang
  Xu.
\newblock Cmt: Convolutional neural networks meet vision transformers.
\newblock In {\em Proceedings of the IEEE/CVF Conference on Computer Vision and
  Pattern Recognition}, pages 12175--12185, 2022.

\bibitem{li2022next}
Jiashi Li, Xin Xia, Wei Li, Huixia Li, Xing Wang, Xuefeng Xiao, Rui Wang, Min
  Zheng, and Xin Pan.
\newblock Next-vit: Next generation vision transformer for efficient deployment
  in realistic industrial scenarios.
\newblock {\em arXiv preprint arXiv:2207.05501}, 2022.

\bibitem{yao2022transclaw}
Chang Yao, Menghan Hu, Qingli Li, Guangtao Zhai, and Xiao-Ping Zhang.
\newblock Transclaw u-net: claw u-net with transformers for medical image
  segmentation.
\newblock In {\em 2022 5th International Conference on Information
  Communication and Signal Processing (ICICSP)}, pages 280--284. IEEE, 2022.

\bibitem{ji2021multi}
Yuanfeng Ji, Ruimao Zhang, Huijie Wang, Zhen Li, Lingyun Wu, Shaoting Zhang,
  and Ping Luo.
\newblock Multi-compound transformer for accurate biomedical image
  segmentation.
\newblock In {\em Medical Image Computing and Computer Assisted
  Intervention--MICCAI 2021: 24th International Conference, Strasbourg, France,
  September 27--October 1, 2021, Proceedings, Part I 24}, pages 326--336.
  Springer, 2021.

\bibitem{zhou2021nnformer}
Hong-Yu Zhou, Jiansen Guo, Yinghao Zhang, Lequan Yu, Liansheng Wang, and Yizhou
  Yu.
\newblock nnformer: Interleaved transformer for volumetric segmentation.
\newblock {\em arXiv preprint arXiv:2109.03201}, 2021.

\bibitem{tu2022maxvit}
Zhengzhong Tu, Hossein Talebi, Han Zhang, Feng Yang, Peyman Milanfar, Alan
  Bovik, and Yinxiao Li.
\newblock Maxvit: Multi-axis vision transformer.
\newblock In {\em Computer Vision--ECCV 2022: 17th European Conference, Tel
  Aviv, Israel, October 23--27, 2022, Proceedings, Part XXIV}, pages 459--479.
  Springer, 2022.

\bibitem{yang2006nuclei}
Xiaodong Yang, Houqiang Li, and Xiaobo Zhou.
\newblock Nuclei segmentation using marker-controlled watershed, tracking using
  mean-shift, and kalman filter in time-lapse microscopy.
\newblock {\em IEEE Transactions on Circuits and Systems I: Regular Papers},
  53(11):2405--2414, 2006.

\bibitem{veta2013automatic}
Mitko Veta, Paul~J Van~Diest, Robert Kornegoor, Andr{\'e} Huisman, Max~A
  Viergever, and Josien~PW Pluim.
\newblock Automatic nuclei segmentation in h\&e stained breast cancer
  histopathology images.
\newblock {\em PloS one}, 8(7):e70221, 2013.

\bibitem{tsai2003shape}
Andy Tsai, Anthony Yezzi, William Wells, Clare Tempany, Dewey Tucker, Ayres
  Fan, W~Eric Grimson, and Alan Willsky.
\newblock A shape-based approach to the segmentation of medical imagery using
  level sets.
\newblock {\em IEEE transactions on medical imaging}, 22(2):137--154, 2003.

\bibitem{held1997markov}
Karsten Held, E~Rota Kops, Bernd~J Krause, William~M Wells, Ron Kikinis, and
  H-W Muller-Gartner.
\newblock Markov random field segmentation of brain mr images.
\newblock {\em IEEE transactions on medical imaging}, 16(6):878--886, 1997.

\bibitem{fu2018joint}
Huazhu Fu, Jun Cheng, Yanwu Xu, Damon Wing~Kee Wong, Jiang Liu, and Xiaochun
  Cao.
\newblock Joint optic disc and cup segmentation based on multi-label deep
  network and polar transformation.
\newblock {\em IEEE transactions on medical imaging}, 37(7):1597--1605, 2018.

\bibitem{kiran2022denseres}
Iqra Kiran, Basit Raza, Areesha Ijaz, and Muazzam~A Khan.
\newblock Denseres-unet: Segmentation of overlapped/clustered nuclei from multi
  organ histopathology images.
\newblock {\em Computers in Biology and Medicine}, 143:105267, 2022.

\bibitem{singha2023alexsegnet}
Anu Singha and Mrinal~Kanti Bhowmik.
\newblock Alexsegnet: an accurate nuclei segmentation deep learning model in
  microscopic images for diagnosis of cancer.
\newblock {\em Multimedia Tools and Applications}, 82(13):20431--20452, 2023.

\bibitem{gu2020net}
Ran Gu, Guotai Wang, Tao Song, Rui Huang, Michael Aertsen, Jan Deprest,
  S{\'e}bastien Ourselin, Tom Vercauteren, and Shaoting Zhang.
\newblock Ca-net: Comprehensive attention convolutional neural networks for
  explainable medical image segmentation.
\newblock {\em IEEE transactions on medical imaging}, 40(2):699--711, 2020.

\bibitem{wang2019sclerasegnet}
Caiyong Wang, Yunlong Wang, Yunfan Liu, Zhaofeng He, Ran He, and Zhenan Sun.
\newblock Sclerasegnet: An attention assisted u-net model for accurate sclera
  segmentation.
\newblock {\em IEEE Transactions on Biometrics, Behavior, and Identity
  Science}, 2(1):40--54, 2019.

\bibitem{lal2021nucleisegnet}
Shyam Lal, Devikalyan Das, Kumar Alabhya, Anirudh Kanfade, Aman Kumar, and
  Jyoti Kini.
\newblock Nucleisegnet: robust deep learning architecture for the nuclei
  segmentation of liver cancer histopathology images.
\newblock {\em Computers in Biology and Medicine}, 128:104075, 2021.

\bibitem{shi2022fine}
Tangqi Shi, Chaoqun Li, Dou Xu, and Xiayue Fan.
\newblock Fine-grained histopathological cell segmentation through residual
  attention with prior embedding.
\newblock {\em Multimedia Tools and Applications}, 81(5):6497--6511, 2022.

\bibitem{milletari2016v}
Fausto Milletari, Nassir Navab, and Seyed-Ahmad Ahmadi.
\newblock V-net: Fully convolutional neural networks for volumetric medical
  image segmentation.
\newblock In {\em 2016 fourth international conference on 3D vision (3DV)},
  pages 565--571. Ieee, 2016.

\bibitem{ali2023cb}
Momina~Liaqat Ali, Zunaira Rauf, Asifullah Khan, Anabia Sohail, Rafi Ullah, and
  Jeonghwan Gwak.
\newblock Cb-hvtnet: A channel-boosted hybrid vision transformer network for
  lymphocyte assessment in histopathological images.
\newblock {\em arXiv preprint arXiv:2305.09211}, 2023.

\bibitem{karimi2021convolution}
Davood Karimi, Serge~Didenko Vasylechko, and Ali Gholipour.
\newblock Convolution-free medical image segmentation using transformers.
\newblock In {\em Medical Image Computing and Computer Assisted
  Intervention--MICCAI 2021: 24th International Conference, Strasbourg, France,
  September 27--October 1, 2021, Proceedings, Part I 24}, pages 78--88.
  Springer, 2021.

\bibitem{sandler2018mobilenetv2}
Mark Sandler, Andrew Howard, Menglong Zhu, Andrey Zhmoginov, and Liang-Chieh
  Chen.
\newblock Mobilenetv2: Inverted residuals and linear bottlenecks.
\newblock In {\em Proceedings of the IEEE conference on computer vision and
  pattern recognition}, pages 4510--4520, 2018.

\bibitem{xiao2021early}
Tete Xiao, Mannat Singh, Eric Mintun, Trevor Darrell, Piotr Doll{\'a}r, and
  Ross Girshick.
\newblock Early convolutions help transformers see better.
\newblock {\em Advances in Neural Information Processing Systems},
  34:30392--30400, 2021.

\bibitem{hu2018squeeze}
Jie Hu, Li~Shen, and Gang Sun.
\newblock Squeeze-and-excitation networks.
\newblock In {\em Proceedings of the IEEE conference on computer vision and
  pattern recognition}, pages 7132--7141, 2018.

\bibitem{chu2023conditional}
Xiangxiang Chu, Zhi Tian, Bo~Zhang, Xinlong Wang, and Chunhua Shen.
\newblock Conditional positional encodings for vision transformers.
\newblock In {\em The Eleventh International Conference on Learning
  Representations}, 2023.

\bibitem{ba2016layer}
Jimmy~Lei Ba, Jamie~Ryan Kiros, and Geoffrey~E Hinton.
\newblock Layer normalization.
\newblock {\em arXiv preprint arXiv:1607.06450}, 2016.

\bibitem{ioffe2015batch}
Sergey Ioffe and Christian Szegedy.
\newblock Batch normalization: Accelerating deep network training by reducing
  internal covariate shift.
\newblock In {\em International conference on machine learning}, pages
  448--456. pmlr, 2015.

\bibitem{hendrycks2016gaussian}
Dan Hendrycks and Kevin Gimpel.
\newblock Gaussian error linear units (gelus).
\newblock {\em arXiv preprint arXiv:1606.08415}, 2016.

\bibitem{vaswani2017attention}
Ashish Vaswani, Noam Shazeer, Niki Parmar, Jakob Uszkoreit, Llion Jones,
  Aidan~N Gomez, {\L}ukasz Kaiser, and Illia Polosukhin.
\newblock Attention is all you need.
\newblock {\em Advances in neural information processing systems}, 30, 2017.

\bibitem{tu2022maxim}
Zhengzhong Tu, Hossein Talebi, Han Zhang, Feng Yang, Peyman Milanfar, Alan
  Bovik, and Yinxiao Li.
\newblock Maxim: Multi-axis mlp for image processing.
\newblock In {\em Proceedings of the IEEE/CVF Conference on Computer Vision and
  Pattern Recognition}, pages 5769--5780, 2022.

\bibitem{zhao2021improved}
Long Zhao, Zizhao Zhang, Ting Chen, Dimitris Metaxas, and Han Zhang.
\newblock Improved transformer for high-resolution gans.
\newblock {\em Advances in Neural Information Processing Systems},
  34:18367--18380, 2021.

\bibitem{han2021transformer}
Kai Han, An~Xiao, Enhua Wu, Jianyuan Guo, Chunjing Xu, and Yunhe Wang.
\newblock Transformer in transformer.
\newblock {\em Advances in Neural Information Processing Systems},
  34:15908--15919, 2021.

\bibitem{dai2021coatnet}
Zihang Dai, Hanxiao Liu, Quoc~V Le, and Mingxing Tan.
\newblock Coatnet: Marrying convolution and attention for all data sizes.
\newblock {\em Advances in Neural Information Processing Systems},
  34:3965--3977, 2021.

\bibitem{shaw2018self}
Peter Shaw, Jakob Uszkoreit, and Ashish Vaswani.
\newblock Self-attention with relative position representations.
\newblock {\em arXiv preprint arXiv:1803.02155}, 2018.

\bibitem{jiang2021transgan}
Yifan Jiang, Shiyu Chang, and Zhangyang Wang.
\newblock Transgan: Two pure transformers can make one strong gan, and that can
  scale up.
\newblock {\em Advances in Neural Information Processing Systems},
  34:14745--14758, 2021.

\bibitem{misra2019mish}
Diganta Misra.
\newblock Mish: A self regularized non-monotonic activation function.
\newblock {\em arXiv preprint arXiv:1908.08681}, 2019.

\bibitem{verma2021monusac2020}
Ruchika Verma, Neeraj Kumar, Abhijeet Patil, Nikhil~Cherian Kurian, Swapnil
  Rane, Simon Graham, Quoc~Dang Vu, Mieke Zwager, Shan E~Ahmed Raza, Nasir
  Rajpoot, et~al.
\newblock Monusac2020: A multi-organ nuclei segmentation and classification
  challenge.
\newblock {\em IEEE Transactions on Medical Imaging}, 40(12):3413--3423, 2021.

\bibitem{tcga}
TCGA.
\newblock Network data.
\newblock \url{http://cancergenome.nih.gov/}, 2006.

\bibitem{contributors2020openmmlab}
MMSegmentation Contributors.
\newblock Openmmlab semantic segmentation toolbox and benchmark, 2020.

\bibitem{pytorch}
Adam Paszke, Sam Gross, Francisco Massa, Adam Lerer, James Bradbury, Gregory
  Chanan, Trevor Killeen, Zeming Lin, Natalia Gimelshein, Luca Antiga, Alban
  Desmaison, Andreas Kopf, Edward Yang, Zachary DeVito, Martin Raison, Alykhan
  Tejani, Sasank Chilamkurthy, Benoit Steiner, Lu~Fang, Junjie Bai, and Soumith
  Chintala.
\newblock Pytorch: An imperative style, high-performance deep learning library.
\newblock In {\em Advances in Neural Information Processing Systems 32}, pages
  8024--8035. Curran Associates, Inc., 2019.

\bibitem{kingma2014adam}
Diederik~P Kingma and Jimmy Ba.
\newblock Adam: A method for stochastic optimization.
\newblock {\em arXiv preprint arXiv:1412.6980}, 2014.

\bibitem{zhuang2020adabelief}
Juntang Zhuang, Tommy Tang, Yifan Ding, Sekhar~C Tatikonda, Nicha Dvornek,
  Xenophon Papademetris, and James Duncan.
\newblock Adabelief optimizer: Adapting stepsizes by the belief in observed
  gradients.
\newblock {\em Advances in neural information processing systems},
  33:18795--18806, 2020.

\bibitem{loshchilov2018decoupled}
Ilya Loshchilov and Frank Hutter.
\newblock Decoupled weight decay regularization.
\newblock In {\em International Conference on Learning Representations}, 2019.

\bibitem{xiang2020bio}
Tiange Xiang, Chaoyi Zhang, Dongnan Liu, Yang Song, Heng Huang, and Weidong
  Cai.
\newblock Bio-net: learning recurrent bi-directional connections for
  encoder-decoder architecture.
\newblock In {\em Medical Image Computing and Computer Assisted
  Intervention--MICCAI 2020: 23rd International Conference, Lima, Peru, October
  4--8, 2020, Proceedings, Part I 23}, pages 74--84. Springer, 2020.

\bibitem{zhan2023fsa}
Bangcheng Zhan, Enmin Song, and Hong Liu.
\newblock Fsa-net: Rethinking the attention mechanisms in medical image
  segmentation from releasing global suppressed information.
\newblock {\em Computers in Biology and Medicine}, page 106932, 2023.

\bibitem{qiao2023mbutransnet}
JunBo Qiao, Xing Wang, Ji~Chen, and MingTao Liu.
\newblock Mbutransnet: multi-branch u-shaped network fusion transformer
  architecture for medical image segmentation.
\newblock {\em International Journal of Computer Assisted Radiology and
  Surgery}, pages 1--8, 2023.

\bibitem{chanchal2022deep}
Amit~Kumar Chanchal, Shyam Lal, and Jyoti Kini.
\newblock Deep structured residual encoder-decoder network with a novel loss
  function for nuclei segmentation of kidney and breast histopathology images.
\newblock {\em Multimedia Tools and Applications}, 81(7):9201--9224, 2022.

\bibitem{wang2023improved}
Ji~Wang, Lulu Qin, Dan Chen, Juan Wang, Bo-Wei Han, Zexuan Zhu, and Guangdong
  Qiao.
\newblock An improved hover-net for nuclear segmentation and classification in
  histopathology images.
\newblock {\em Neural Computing and Applications}, pages 1--15, 2023.

\bibitem{vo2023mulvernet}
Vi~Thi-Tuong Vo and Soo-Hyung Kim.
\newblock Mulvernet: Nucleus segmentation and classification of pathology
  images using the hover-net and multiple filter units.
\newblock {\em Electronics}, 12(2):355, 2023.

\bibitem{liu2020scam}
Zuhao Liu, Huan Wang, Shaoting Zhang, Guotai Wang, and Jin Qi.
\newblock Nas-scam: Neural architecture search-based spatial and channel joint
  attention module for nuclei semantic segmentation and classification.
\newblock In {\em Medical Image Computing and Computer Assisted
  Intervention--MICCAI 2020: 23rd International Conference, Lima, Peru, October
  4--8, 2020, Proceedings, Part I 23}, pages 263--272. Springer, 2020.

\bibitem{zhao2017pyramid}
Hengshuang Zhao, Jianping Shi, Xiaojuan Qi, Xiaogang Wang, and Jiaya Jia.
\newblock Pyramid scene parsing network.
\newblock In {\em Proceedings of the IEEE conference on computer vision and
  pattern recognition}, pages 2881--2890, 2017.

\bibitem{xiao2018unified}
Tete Xiao, Yingcheng Liu, Bolei Zhou, Yuning Jiang, and Jian Sun.
\newblock Unified perceptual parsing for scene understanding.
\newblock In {\em Proceedings of the European conference on computer vision
  (ECCV)}, pages 418--434, 2018.

\end{thebibliography}

\end{document}